\documentclass[aps,prb,floats]{revtex4}
\usepackage{color,ulem,verbatim,float,wrapfig}
\usepackage{amssymb,amsbsy,amsmath,mathrsfs}
\usepackage{epsfig,subfigure}
\usepackage{graphicx}
\usepackage{times}
\usepackage{color}
\usepackage{subfigure}
\usepackage{setspace}
\usepackage{bm} 
\usepackage{caption}
\newcommand\bea{\begin{eqnarray}}
\newcommand\eea{\end{eqnarray}}
\newcommand\beq{\begin{equation}}
\newcommand\eeq{\end{equation}}
\newcommand\bib{\bibitem}

\newcommand{\non}{\nonumber}
\newcommand{\al}{\alpha}
\newcommand{\de}{\delta}

\newcommand{\ep}{\epsilon}

\newcommand{\lm}{\lambda}
\newcommand{\si}{\sigma}
\newcommand{\ta}{\theta}
\newcommand{\om}{\omega}
\newcommand{\da}{\dagger}
\newcommand{\pa}{\partial}

\begin{document}


\title{Transport on a topological insulator surface with a time-dependent
magnetic barrier}

\author{Adithi Udupa$^1$, K. Sengupta$^2$ and Diptiman Sen$^{1,3}$}

\affiliation{$^1$Center for High Energy Physics and $^3$Department of Physics,
Indian Institute of Science, Bengaluru 560012, India \\
$^2$School of Physical Sciences, Indian Association for the Cultivation of 
Science, Jadavpur, Kolkata 700032, India}

\begin{abstract}
We study transport across a time-dependent magnetic barrier present on 
the surface of a three-dimensional topological insulator. We show that 
such a barrier can be implemented for Dirac electrons on the surface of a 
three-dimensional topological insulator by a combination of a proximate 
magnetic material and linearly polarized external radiation. We find that 
the conductance of the system can be tuned by varying the frequency and 
amplitude of the radiation and the energy of an electron incident on the 
barrier providing us optical control on the conductance of such
junctions. We first study a $\de$-function barrier which shows a
number of interesting features such as sharp peaks and dips in the
transmission at certain angles of incidence. Approximate methods for
studying the limits of small and large frequencies are presented. We
then study a barrier with a finite width. This gives rise to some
new features which are not present for a $\de$-function barrier,
such as resonances in the conductance at certain values of the
system parameters. We present a perturbation theory for studying the
limit of large driving amplitude and use this to understand the
resonances. Finally, we use a semiclassical approach to study
transmission across a time-dependent barrier and show how this can
qualitatively explain some of the results found in the earlier
analysis. We discuss experiments which can test our theory.
\end{abstract}

\maketitle

\section{Introduction}
\label{sec1}

Topological insulators (TI) have been studied extensively both
theoretically and experimentally due to their remarkable physical
and mathematical properties (see Refs.~[\onlinecite{hasan,qi}] for
reviews). These are materials in which the bulk states are separated
from the Fermi energy by a finite gap and therefore do not
contribute to electronic transport at low temperatures; however,
there are states at the boundaries which are gapless (if certain
symmetries like time-reversal are not broken), and they participate
in transport. In reality, many TIs have some bulk conductance
because the Fermi energy lies within a bulk conduction band.
However, an appropriate amount of doping can produce ideal TIs where
the bulk conductance is very small~\cite{brahlek,kushwaha}. Further,
the number of boundary states with a given momentum is given by a
topological invariant which characterizes the bulk states. In
three-dimensional (3D) topological insulators such as Bi$_2$Se$_3$
and Bi$_2$Te$_3$, the boundaries are surfaces, and the surface
electrons are typically governed by the Hamiltonian of a single
massless Dirac particle in two spatial dimensions. Further, the
Hamiltonians exhibit spin-momentum locking, so that the linear
momentum and the spin angular momentum of an electron are
perpendicular to each other. If time-reversal symmetry is not broken
(by, say, magnetic impurities), the spin-momentum locking leads to
ballistic transport; this is because scattering from a non-magnetic
impurities cannot flip the spin of an electron and therefore cannot
change its momentum. The application of an in-plane magnetic field
which has a Zeeman coupling to the spin or, equivalently, the
presence of a proximate magnetic material which induces in plane
Zeeman magnetization, alters the spin-momentum locking of the Dirac
electrons~\cite{yoko,sm1}. It has been shown in Ref.\ \onlinecite{sm1} 
that the presence of such proximate magnetic material over a strip of 
finite width can cut off conductance across it for sufficiently strong 
induced magnetization. This allows one to implement a magnetic switch 
and achieve magnetic control over electric current in these materials 
utilizing spin-momentum locking of Dirac quasiparticles.

The effects of periodic driving of quantum many-body systems is another
subject which has been studied extensively for about a decade from many points
of view (for reviews, see Refs.~[\onlinecite{rudner,nathan,cayssol,goldman,
eckardt,bukov,mikami}]). Such periodic driving can be used to change the band
structure of a material (giving rise to phenomena such as dynamical
freezing~\cite{das,bhat,hegde,pekker,nag1,nag2,agarwala1}), generate
boundary modes and induce dynamical topological transitions by changing a
non-topological system to a topological one~\cite{rudner,cayssol,mikami,oka,
inoue,kita1,kita2,gu,lindner,morrell,katan,delplace,tong,thakur1,thakur2,usaj,
kundu1,carp,alessio,xiong,loss,mukh1,mukh2,zhou}, produce novel steady states
which cannot appear in time-independent systems~\cite{nathan}, and control
electronic transport~\cite{bukov,kundu2,agarwala2}.

\begin{figure}[H]
\centering
\hspace*{-.4cm} \includegraphics[scale=0.45]{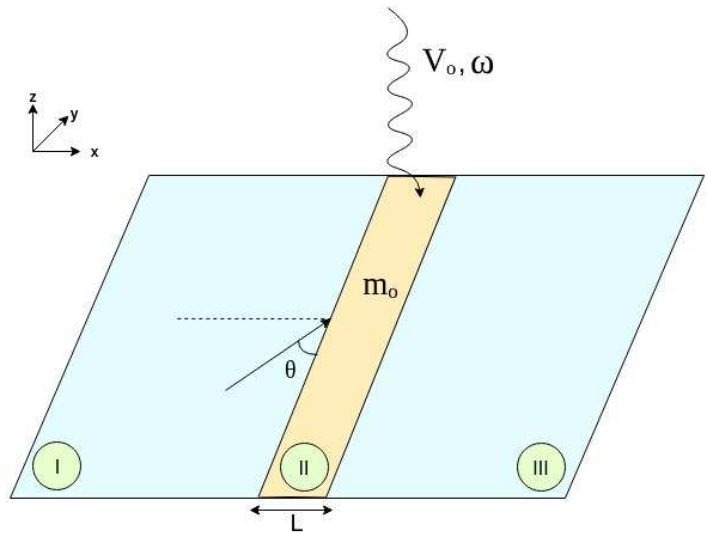}
\caption{Schematic picture of the system showing the top surface of a TI,
with a static magnetic barrier of strength $m_0$ and electromagnetic
radiation with frequency $\om$ and amplitude $V_0$ which is incident on the
barrier. The barrier has a width $L$ (region $II$), and an electron comes in
from region $I$ with an angle of incidence $\ta$.} \label{fig01} \end{figure}

In this paper, we will be interested in the last aspect of periodic
driving, namely its role on the electric transport. More precisely,
we will study what happens when a magnetic barrier which is
periodically varying in time is placed on the surface of a 3D TI
like Bi$_2$Se$_3$. Such a barrier can be realized in these materials
by applying an in-plane magnetic field (which induces a static
Zeeman magnetization) and a linearly polarized light (which provides
the time varying part) over a region of width $L$. A schematic
picture of the system is shown in Fig.~\ref{fig01}. We will see that
the conductance of this system, in the presence of such a barrier,
exhibits a number of interesting features, such as prominent peaks
and dips, as the different system parameters are varied. Moreover, such
a barrier provides a route to optical (electromagnetic) control over
electric current in such junctions; we show that the conductance of such a
junction can be tuned by controlling the amplitude and frequency of the 
applied light even in the absence of a static Zeeman field. We note that 
an analogous study for the case of an oscillating potential barrier was 
carried out in Ref.~\onlinecite{mondal}; however, the optical control of 
conductance that we find in our study has not been obtained earlier.

The plan of this paper is as follows. In Sec.~\ref{sec2} we study
the effect of a $\de$-function magnetic barrier which has both a
constant term $m_0$ and a term which oscillates sinusoidally with an
amplitude $\al$ and frequency $\om$; we take the field to point
along the $\hat x$ direction. The $\de$-function nature of the
barrier allows a simpler analysis of the problem than the more
realistic case of a finite-width barrier which is discussed in the
next section. We find that the $\de$-function leads to a non-trivial
matching condition for the wave function on the left and right edges
of the barrier. Assuming that an electron is incident on the barrier
from the left with an energy $E_0$ and angle of incidence $\ta$, we
will use the matching condition to derive the transmission
amplitudes in the different Floquet bands and hence the transmitted
current on the right of the barrier. Integrating this over $\ta$
gives the differential conductance $G$. We then present our
numerical results for the transmitted current and $G$ as a function
of the parameters $m_0$, $\al$, $\om$ and $E_0$. Several interesting
features are seen, such as kinks in the transmitted current at
certain values of $\ta$ for intermediate values of $\om$ and peaks
in $G$ at certain values of $m_0$ for small values of $\om$. We
provide simple explanations for these features and for the results
obtained in some limits of the problem such as small and large
values of $\om$. In Sec.~\ref{sec3}, we study a time-dependent
magnetic barrier which has a finite width $L$; the field is taken to
consist of a constant $m_0$ and an oscillating term with an
amplitude $V_0$ and frequency $\om$. We examine surface plots of $G$
as a function of $V_0$ and $m_0$ for various values of $\om$. We
find that $G$ is peaked along the line $m_0 = V_0$ for small $\om$
and shows resonance-like features for a discrete set of values of
$V_0$ when $\om$ is large and $m_0$ is small. We show that the
conductance of a time-dependent barrier with parameters
$(m_0,V_0,\om)$ can be mapped to that of a time-independent barrier
with a single parameter $m_{eff}$; we find that $m_{eff}$ approaches
a constant for large $\om$, where the constant depends on
$(m_0,V_0)$. We also find that when $E_0$ and $\om$ are large and
are related as $E_0 = \hbar \om/2$, there are prominent dips in $G$.
We analyze how the width and magnitude of the dips depend on the
parameters $V_0$ and $L$. We provide a detailed study of the
behavior of $G$ as a function of $m_0$ and $V_0$ at both high and
low frequencies, and show that for small or zero $m_0$, one can tune
$V_0$ to control $G$ in these systems. This shows that these
junctions can operate as optically controlled switches. To study
this more carefully, we carry out a detailed investigation of the
dependence of $G$ on $V_0$ and $\om$, for a fixed value of $E_0$ and
$L$, taking $m_0 = 0$. We find that for $\om / V_0 \gtrsim 0.83$,
there are curves in the $(V_0,\om)$ plane along which there are
resonances and $G$ is particularly high; the spacing between these
curves is proportional to $1/L$. For $\om / V_0 \lesssim 0.83$, the
conductance is small everywhere; however, it is particularly small
along certain straight lines. We present a Floquet perturbation
theory which can explain both the resonances and the lines of very
small conductances. We summarize our results and discuss possible
experimental tests in Sec.~\ref{sec5}. The paper ends with two
appendices. In Appendix A we briefly recall the basics of Floquet
theory and how the Floquet eigenstates and eigenvalues can be found.
In Appendix B, we present a completely different approach to the
problem of a time-dependent magnetic barrier. While Secs.~\ref{sec2}
and \ref{sec3} considered a spin-1/2 electron, we consider the
large-spin limit in this appendix and study the semiclassical
equations of motion for the trajectory of such a particle moving in
two dimensions in the presence of an electromagnetic field. Using this 
approach, we are able to qualitatively understand some features of the 
surface plots of $G$ versus $(V_0,m_0)$ presented in Sec.~\ref{sec3}.

\section{$\de$-function magnetic barrier}
\label{sec2}

In this section, we will study the effect of a magnetic barrier present
on the top surface of a three-dimensional TI~\cite{hasan,qi}. The
Hamiltonian for electrons on the surface (taken to be the $x-y$
plane) is known to have the form of a massless Dirac equation with
spin-momentum locking, namely,
\beq H ~=~ -i \hbar v ~\left[\si^x \frac{\pa}{\pa y} ~-~ \si^y
\frac{\pa}{\pa x} \right], \label{ham1} \eeq
where $v$ is the velocity, $\si^a$ denote Pauli matrices, and
the wave function $\psi (x,y,t)$ has two components. In the well-known 
TI Bi$_2$Se$_3$, it is known~\cite{qi} that $\hbar v = 0.333$ eV-nm.

We now consider a magnetic barrier which has the form of a
$\de$-function located along the line $x=0$; the barrier strength
will be taken to have both a constant term and a term which
oscillates in time with a frequency $\om$ and an amplitude $\alpha$.
We will assume that the magnetic field associated with the barrier
points in the $\hat{x}$ direction; we can then ignore the coupling
of the field to the orbital motion of the electrons since they are
constrained to move in the $x-y$ plane and therefore cannot have
cyclotron orbits perpendicular to the field direction. Hence we only
have a Zeeman coupling of the field to the electron spin. The
Hamiltonian will therefore take the form \beq H ~=~ \hbar v ~\left[
\si^x \{-i \frac{\pa}{\pa y} ~+~ [m_0 ~+~ \al ~ \cos (\om t)] ~\de
(x)\} ~+~ i \si^y \frac{\pa}{\pa x} \right], \label{ham2} \eeq where
for an applied field $B= B_0 + B_1 \cos(\omega t) \hat x$, $m_0
[\alpha]= g \mu_B B_0[B_1]/(\hbar v k_0)$, $g$ is the gyromagnetic
ratio, $\mu_B$ is the Bohr magneton, and $k_0$ is the inverse of a
typical length scale (for instance, a typical barrier width of the
order of 20 nm as stated in Sec.~\ref{sec3a}). We note that $m_0$
and $\al$ are dimensionless quantities. We also point out that the
oscillation term may be generated via coupling of $H$ to linearly
polarized vector potential given by $A = A_0 \cos(\omega t) \hat y$
in which case $\alpha= e A_0/(\hbar v k_0)$, where $e$ is the
electron charge. This is a consequence of the fact that the coupling
to the magnetic barrier appears in Eq.~\eqref{ham2} as an addition
to $p_y = -i \hbar \pa / \pa y$ and therefore acts as a vector
potential in the $y$ direction.

Given that the wave function satisfies
\beq i \hbar \frac{\pa \psi (x,y,t)}{\pa t} ~=~ H \psi (x,y,t),
\label{sch} \eeq
we can derive the matching condition that $\psi$ must satisfy at $x=0$ due
to the $\de$-function in Eq.~\eqref{ham2}. To do this, we assume that
\beq \psi ~=~ \psi_0 (y,t) ~e^{f(x)} \eeq
in the vicinity of $x=0$, where $f(x)$ may have a discontinuity at $x=0$.
We then integrate the two sides of Eq.~\eqref{sch} from $x=-\ep$ to $x=+\ep$,
and take the limit $\ep \to 0$. This gives
\beq i \si^y ~[f(x=0+) ~-~ f(x=0-)] ~+~ \si^x ~[m_0 ~+~ \al \cos (\om t)]
~=~ 0, \eeq
which implies that
\beq f(x=0+) ~-~ f(x=0-) ~=~ \si^z ~[m_0 ~+~ \al \cos (\om t)]. \eeq
Thus the wave function will have a discontinuity at $x=0$ given by
\bea \psi_{x=0^+} &=& e^{[m_0 + \al ~ \cos (\om t)] \si^z}
\psi_{x=0^-}, \non \\
{\rm namely,} ~~~\psi_{x=0^+} &=& \begin{pmatrix}
e^{m_0 + \al \cos (\om t)} & 0 \vspace{0.5cm} \\
0 & e^{- m_0 - \al \cos (\om t)} \end{pmatrix} \psi_{x=0^-}.
\label{delcon}\eea

We can now use the identity
\beq e^{\al \cos \ta} ~=~ \sum_{n=-\infty}^\infty ~I_n(\al) ~e^{-in\ta}, \eeq
where $I_n(\al)$ is the modified Bessel function~\cite{abram}. (These functions
satisfy $I_n (\al) = I_{-n} (\al)$ and $I_n (-\al) = (-1)^n I_n (\al)$).
Hence Eq.~\eqref{delcon} takes the form
\beq \psi_{x = 0^+} ~=~ \begin{pmatrix} e^{m_0}\sum\limits_{n=-\infty}^{
\infty}I_n (\al) e^{-in\om t} & 0 \vspace{0.5cm}\\ 0 & e^{-m_0}
\sum\limits_{n=-\infty}^{\infty}I_n (-\al) e^{-in\om t}
\end{pmatrix} ~\psi_{x= 0^-}. \label{expcond} \eeq

\subsection{Transmitted particle current}
\label{sec2a}

We now assume that an electron is incident on the magnetic barrier from 
the left ($x<0$) with energy $E_0 > 0$ (measured with respect to the Dirac 
point) and momentum $(k_{x,0},k_y)$. (These satisfy the dispersion relation 
$E_0 = v\sqrt{k_{x,0}^2 + k_y^2}$). We will calculate the probabilities of 
reflection and transmission from the time-dependent barrier.

We now use the matching condition in Eq.~\eqref{expcond}.
In our problem, $\psi_{x = 0^+} = \psi_t$ is the transmitted wave, while
$\psi_{x = 0^-}$ is given by the sum of the incident wave $\psi_i$ and the
reflected wave $\psi_r$. The incident wave is given by
\beq \frac{1}{\sqrt{2}} ~\begin{pmatrix}
1 \vspace{0.5cm}\\ \dfrac{E_0}{v ~(k_y+ik_{x,0})} \end{pmatrix}, \eeq
where $k_{x,0}=\sqrt{(E_0 /v)^2- k_y^2}$. This gives rise to reflected and
transmitted waves at $x=0$ due to the presence of the $\de$-function. We can
see from Eq.~\eqref{psit} that the allowed energies of these are given by the
energies of all the Floquet modes, namely,
\beq E_n ~=~ E_0 ~+~ n \hbar \om, \eeq
where $n$ is an integer which can take any value from $-\infty$ to $\infty$;
the modes with $n \ne 0$ are called side bands. (Note that $E_n$ may be 
either positive or negative). Since the $\de$-function is independent of the 
$y$ coordinate, the momentum $k_y$ is a good quantum number. For each energy 
$E_n$, the corresponding eigenfunction is given by
\beq \frac{1}{\sqrt{2}}\begin{pmatrix}
1 \vspace{0.5cm}\\ \dfrac{E_n}{v(k_y+ik_{x,n})} \end{pmatrix} ~~~
\text{with} ~~~ k_{x,n} ~=~ \pm \sqrt{(E_n/v)^2 ~-~ k_y^2}. \label{ef} \eeq
The $\pm$ sign for $k_{x,n}$ in Eq.~\eqref{ef} is fixed by the requirement
that the group velocity $v_g = \pa E_n/\pa k_{x,n}$ should be positive for
the transmitted wave in the region $x > 0$ and negative for the reflected
wave in the region $x < 0$. (This holds if $k_{x,n}$ is real. If $k_{x,n}$
is imaginary, we have to choose the $\pm$ sign in such a way that the
corresponding wave decays as $x \to + \infty$ or $- \infty$). Substituting 
the eigenfunctions in Eq.~\eqref{ef} in Eq.~\eqref{expcond}, we find that 
at $x \to 0$,
\begin{small}
\beq \begin{pmatrix}
e^{m_0} \sum\limits_{m=-\infty}^{\infty}I_m (\al) e^{-im\om t} \hspace{-1cm}&
0 \vspace{0.5cm}\\ 0 & e^{-m_0} \sum \limits_{m=-\infty}^{\infty} I_m (-\al)
e^{-im\om t} \end{pmatrix} \Bigg[ \begin{pmatrix} 1 \vspace{0.5cm}\\ \beta_0^+
\end{pmatrix} ~e^{-iE_0~t} ~+~ r_n \begin{pmatrix}
1 \vspace{0.5cm}\\ \beta_n^-\end{pmatrix} ~e^{-iE_n~t}\Bigg] ~=~ t_n
\begin{pmatrix}
1 \vspace{0.5cm}\\ \beta_n^+ \end{pmatrix} ~e^{-iE_n~t} \label{eqn1}, \eeq
\end{small}
where
\beq \beta_n^{\pm} ~=~ \dfrac{E_n}{v(k_y \pm ik_{x,n})}, \label{betan} \eeq
and $r_n$ and $t_n$ are the reflection and transmission amplitudes
respectively for the $n$-th Floquet mode with energy $E_n$. The superscript
$\pm$ in $\beta_n^\pm$ indicates if the wave is traveling in the positive or
negative $x$ direction.

Given the Hamiltonian in Eq.~\eqref{ham2}, we can use the equation
of continuity to show that the particle current operator is given by
\beq J_x ~=~ - ~v \si^y. \label{jx} \eeq The charge current will be
equal to the particle current multiplied by the electron charge, as
we will see in Eq.~\eqref{condG} below. Using the eigenfunctions in
Eq.~\eqref{ef}, we find that the transmitted particle current in the
band $n$ is given by 
\beq <J_x>_n ~=~ \dfrac{v^2 |t_n|^2 k_{x,n} }{E_n} \eeq 
if $k_{x,n}$ is real. (If $k_{x,n}$ is imaginary, we find that $<J_x>_n = 0$.
This is because when $k_{x,n}$ is imaginary, both the components of the
corresponding wave function are real and hence the expectation value of
$\si^y$ vanishes).
Thus $E_n$ and $k_{x,n}$ must have the same sign in value of $<J_x>$ in the
transmitted region in order to have a positive value of $<J_x>_n$. Similarly,
$E_n$ and $k_{x,n}$ must have opposite signs to have a negative value
of $<J_x>_n$. Now, since $E_n = E_0 + n \hbar \om$, and $e^{-in\om t}$ form
a basis for periodic functions of $t$, we can match the coefficients for each
$n$ separately in Eq.~\eqref{eqn1}. For each value of $n$, we obtain two
equations since the wave function has two components. To solve these equations
numerically, we must truncate the number of equations.
If $N$ is the total number of values that the integer $n$ can take, we have
$2N$ equations, and $2N$ unknown coefficients ($N$ reflection amplitudes $r_n$
and $N$ transmission amplitudes $t_n$). We can therefore numerically find the
values of $r_n$ and $t_n$. Incidentally, current conservation implies
that~\cite{moskalets,agarwal}
\beq \frac{k_{x,0}}{E_0} ~=~ \sum_{n=-\infty}^\infty ~(|t_n|^2 ~+~ |r_n|^2)~
\frac{k_{x,n}}{E_n}. \eeq

Rearranging Eqs.~\eqref{eqn1}, we obtain the matrix equation
\beq \begin{pmatrix}
\ddots & \vdots & \vdots & \vdots & \vdots & \vdots & \vdots & \vdots &
\reflectbox{$\ddots$} \\
\cdots & e^{m_0}I_0^+ & e^{m_0}I_{-1}^+ & e^{m_0}I_{-2}^+ & \cdots & -1 & 0 &
0 & \cdots\\
\cdots & e^{m_0}I_1^+ & e^{m_0}I_0^+ & e^{m_0}I_{-1}^+ & \cdots & 0 & -1 & 0 &
\cdots \\
\cdots & e^{m_0}I_2^+ & e^{m_0}I_1^+ & e^{m_0}I_0^+ & \cdots & 0 & 0 & -1 &
\cdots \\
& \vdots & \vdots & \vdots & \vdots & \vdots & \vdots & \vdots \\
\cdots & e^{-m_0}\beta_{-1}^-I_0^- & e^{-m_0}\beta_0^- I_{-1}^- &
e^{-m_0} \beta_1^- I_{-2}^- & \cdots & \beta_{-1}^+0 & 0 & 0& \cdots \\
\cdots & e^{-m_0}\beta_{-1}^- I_1^- & e^{-m_0}\beta_0^- I_0^- &
e^{-m_0}\beta_1^-I_{-1}^- & \cdots & 0 & \beta_0^+& 0 & \cdots \\
\cdots & e^{-m_0}\beta_{-1}^- I_2^- & e^{-m_0}\beta_0^- I_1^- &
e^{-m_0}\beta_1^- I_0^- & \cdots & 0 & 0 & \beta_1^+ &\cdots \\
\reflectbox{$\ddots$} & \vdots & \vdots & \vdots & \vdots & \vdots & \vdots &
\vdots & \ddots \\
\end{pmatrix}
\begin{pmatrix}
\vdots \\
r_{-1} \\ r_0 \\ r_1 \\ \vdots \\ t_{-1} \\ t_0 \\ t_1 \\ \vdots
\end{pmatrix} =- \begin{pmatrix}
\vdots \\
e^{m_0}I_{-1}^+ \\
e^{m_0}I_0^+\\
e^{m_0}I_1^+\\
\vdots \\
e^{-m_0}\beta_0^+ I_{-1}^- \\
e^{-m_0}\beta_0^+ I_0^-\\
e^{-m_0}\beta_0^+ I_1^-\\
\vdots
\end{pmatrix}. \label{mat} \eeq
where $I^+_n$ and $I^-_n$ denote the $n$-th modified Bessel function with
arguments $\al$ and $-\al$ respectively.

The total transmitted particle current is given by a sum over all bands,
\beq I_t ~=~ <J_x>_{tot} ~=~ \sum\limits_{n=-\infty}^{\infty} ~|t_n|^2 \bigg(
\dfrac{v^2 k_{x,n}}{E_n}\bigg), \label{it} \eeq
where the sum only runs over terms in which $k_{x,n}$ is real.

The angle of incidence of the electron coming from the left of the
barrier is given by \beq \ta ~=~ \tan^{-1} \Big(\frac{k_x}{k_y}
\Bigr), \eeq where $\ta$ lies in the range $[0,\pi]$; normal
incidence corresponds to $\ta = \pi/2$. The total transmitted
particle current $I_t$ is clearly a function of $\ta$. Given $I_t
(\ta)$, the differential conductance $G$ can be calculated as
follows~\cite{adithi}. Suppose that the $\mu_L$ and $\mu_R$ denote
the chemical potentials of the left and right leads assumed to be at
$x \to - \infty$ and $x \to \infty$ respectively; the chemical
potential $\mu$ in a lead is related to the voltage $V$ applied to
that lead as $\mu = qV$, where $q$ is the charge of the electron. In
the zero-bias limit, $\mu_L, ~\mu_R ~\to~ E_0$, the differential
conductance is given by \beq G ~=~ \frac{dI}{dV} ~=~ \frac{q^2 W
E_0}{(2\pi v\hbar)^2} ~\int_0^{\pi} d\ta ~I_t (\ta), \label{condG}
\eeq where $W$ is the width of the system in the $y$ direction
(assumed to be much larger that the wavelength $\sim \hbar v/E_0$ of
the electrons). It is convenient to define a quantity $G_0$ which is
the maximum possible value of $G$; this arises when the total transmitted 
current has the maximum possible value given by $I_t (\ta) = v^2 k_{x,0}/E_0 
= v \sin \ta$. The conductance in this case is given by
\beq G_0 ~=~ \frac{q^2 W E_0}{2v (\pi \hbar)^2}. \label{condG0} \eeq
In the figures presented below, we will plot the dimensionless ratio
$G/G_0 = (1/2v) \int_0^{\pi} d\ta I_t (\ta)$ whose maximum value is 1.

\subsection{Numerical results}
\label{sec2b}

We now present our numerical results as a function of the different parameters 
of the system. In our calculations, we have generally taken $E_0 = 2$
in units of $0.01$ eV (so that it is much smaller than the bulk gap of $0.56$
eV in Bi$_2$Se$_3$, ensuring that there is no contribution to the current from
the bulk states). Hence $k = \sqrt{k_x^2 + k_y^2}$ will have the value
$E_0/(\hbar v)= 0.02/0.333 = 0.06$ nm$^{-1}$ in Bi$_2$Se$_3$. The values of
$\om$ will be taken to be in units of $0.01$ eV/$\hbar$ $\simeq 15.2$ THz.

\subsubsection{Transmitted particle current as a function of $\ta$}
\label{sec2b1}

\begin{figure}[H]
\centering
\subfigure[]{\includegraphics[scale=0.50]{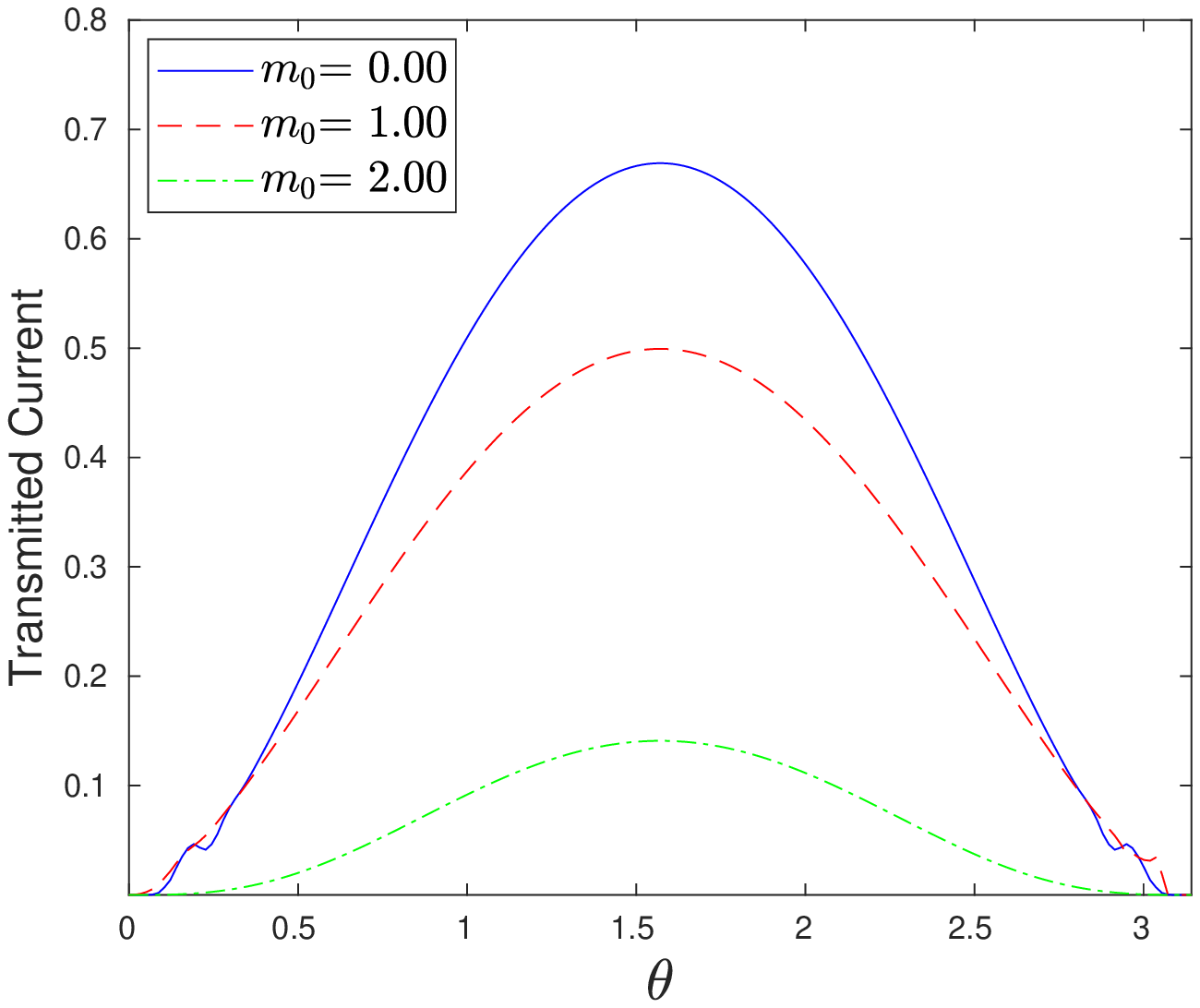}}
\subfigure[]{\includegraphics[scale=0.50]{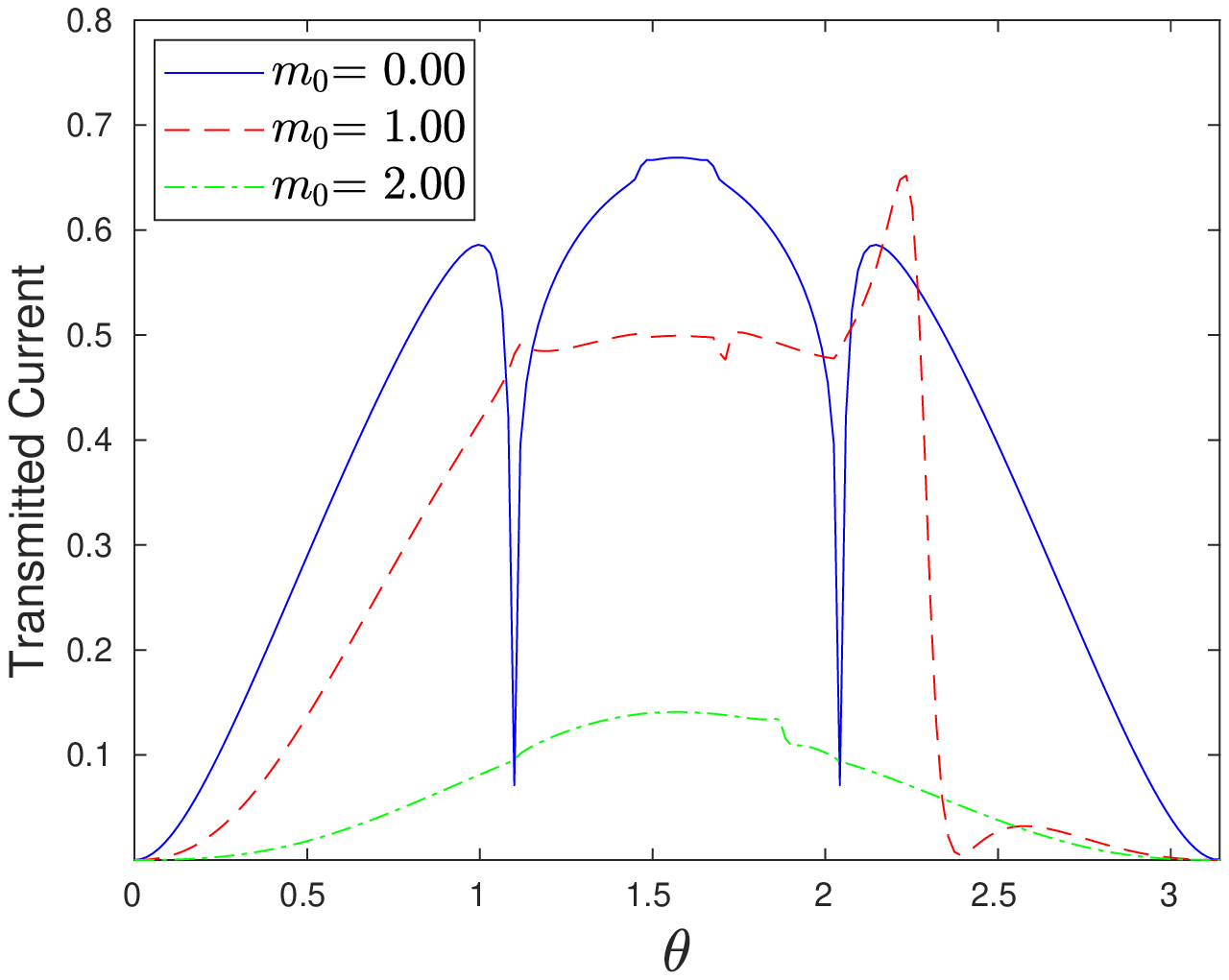}}
\subfigure[]{\includegraphics[scale=0.50]{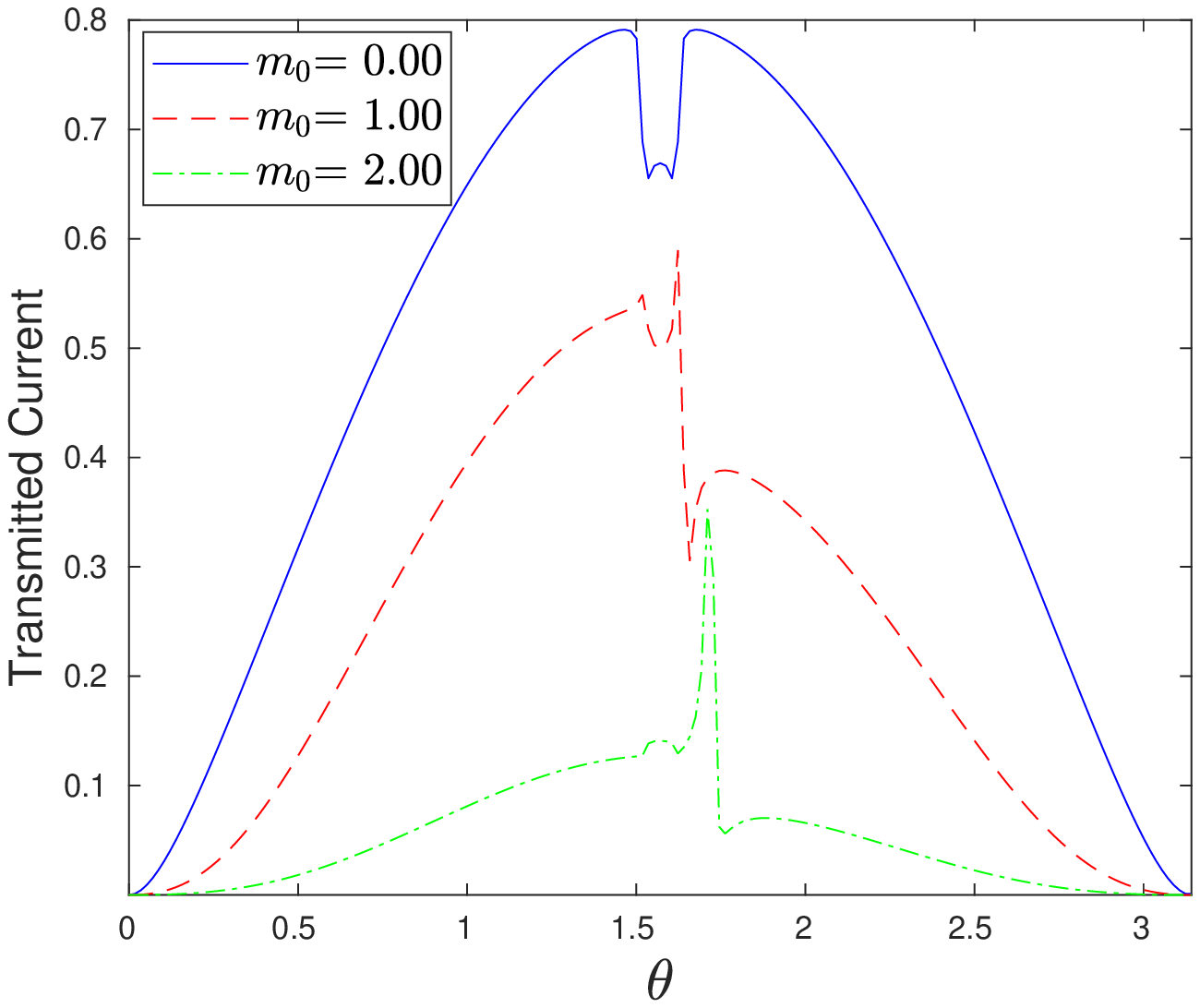}}
\subfigure[]{\includegraphics[scale=0.50]{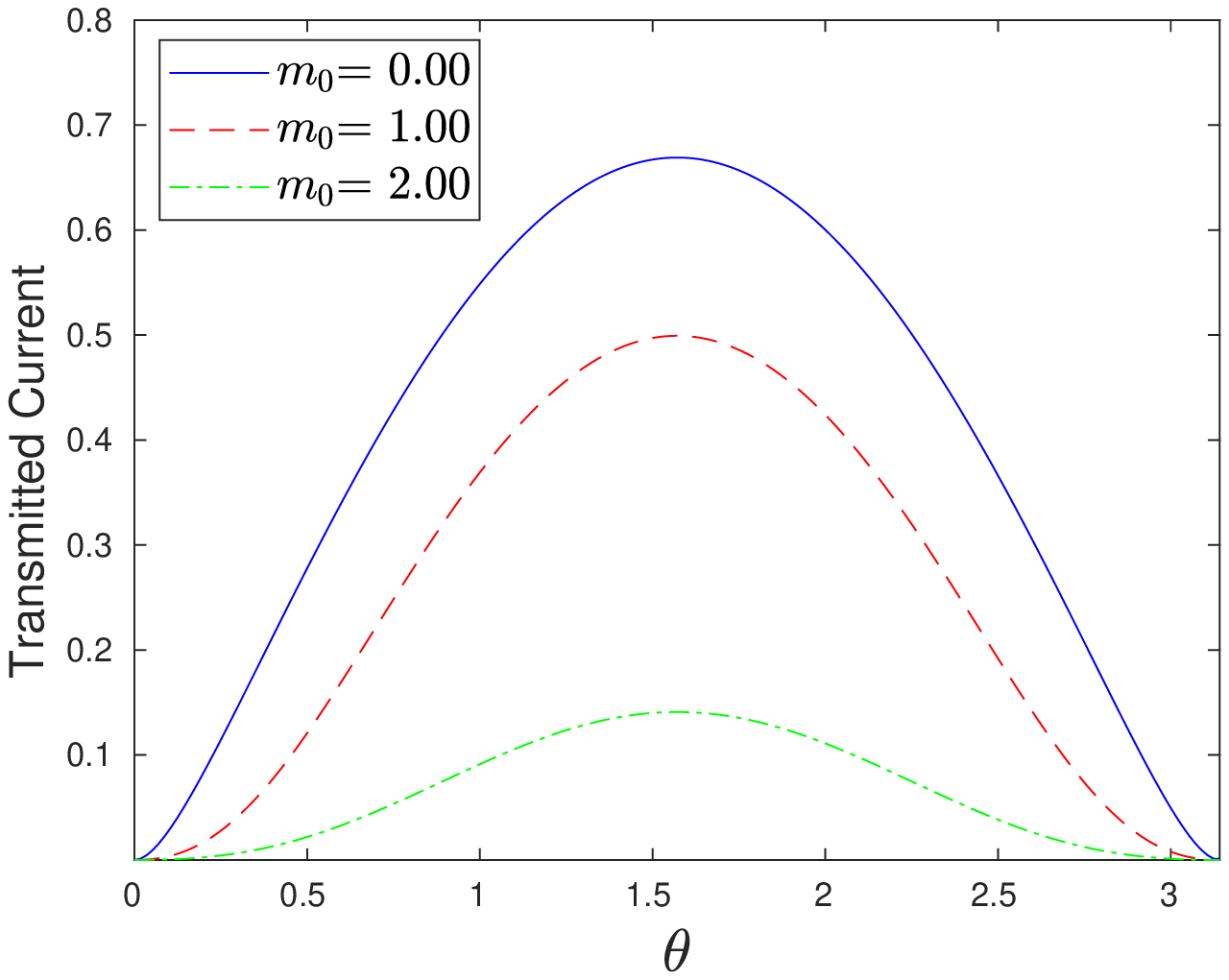}}
\caption{Transmitted particle current as a function of angle of incidence for
different values of $m_0$, $E_0= 2, ~\alpha = 1$, and $\om$ equal to (a) 0.01,
(b) 1.1, (c) 2.1, and (d) 5.1.} \label{fig02} \end{figure}

Figure~\ref{fig02} shows the total transmitted particle current as a
function of the angle of incidence $\ta$, for different values of $m_0$, $\al$
and $\om$. We observe that there are kinks in
some of the plots. These occur because at those values of $\ta$, the value of
$k_{x,n}$ changes from real to imaginary for some value of $n$; as a result
the contribution to the transmitted current from that side band becomes zero.
Given that $v \hbar k_{x,n} = \pm \sqrt{(E_0 + n \hbar \om)^2- v^2 \hbar^2
k_y^2}$ $= \pm \sqrt{(E_0 + n \hbar \om)^2- E_0^2 \cos^2 \ta}$, we see that
$k_{x,n}$ becomes imaginary when
\beq \cos \ta ~=~ \pm ~\left(1 ~+~ \dfrac{n \hbar \om}{E_0} \right).
\label{costa} \eeq
It can be checked that the kinks appearing in Fig.~\ref{fig02}
exactly coincide with the values of $\ta$ corresponding to different values of
$n$ in the above equation. For small and large values of $\om$, kinks do not
appear. For large $\om$, there is no value of $n$ which satisfies the condition
in Eq.~\eqref{costa}. For small $\om$, the values of $\ta$ which satisfy the
condition lie close to glancing angles (0 and $\pi$) where the transmitted
current is always small; hence kinks in the current are not observable.

For $m_0 = 0$, Fig.~\ref{fig02} shows that the transmitted particle current is
symmetric about $\ta = \pi /2$. This is because of the following
symmetry of the Hamiltonian in Eq.~\eqref{ham2} when $m_0 = 0$. Replacing $-i
\pa /\pa y \to k_y = k \cos \ta$, we see that changing $\ta \to \pi - \ta$ and
shifting the time $t \to t + \pi/\om$, we get a new Hamiltonian which is
related to the old Hamiltonian by the unitary transformation $H \to \si^y H
\si^y$. According to Floquet theory, physical quantities like the transmitted
current are invariant under time shifts. Hence the current must be invariant
under $\ta \to \pi - \ta$.

\subsubsection{Differential conductance as a function of $m_0$}
\label{sec2b2}

\begin{figure}[H]
\centering
\includegraphics[scale=0.50]{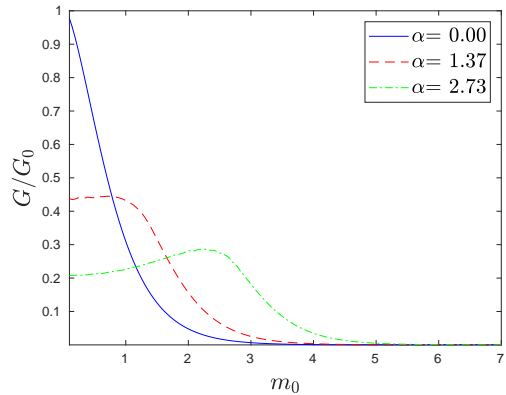}
\caption{$G/G_0$ as a function of $m_0$ for different values of $\al$, with
$E_0= 2$, and $\om$ equal to $0.01$.} \label{fig03} \end{figure}

Figure~\ref{fig03} shows the dimensionless differential conductance $G/G_0$ as
a function of $m_0$, for different values of $\al$. If $\al$ is not too small, 
we see that there are peaks in
the conductance, and their locations move with $\al$. This can be qualitatively
understood as follow, particularly for the lowest value of $\om = 0.01$.
The strength of the magnetic barrier is given by $m_0 + \al \cos (\om t)$.
When $m_0 = \pm \al$, the barrier strength stays close to zero for a long
time since $\pm 1$ correspond to the extreme values of $\cos (\om t)$; this is
particularly true if $\om$ is small. The barrier strength being close to zero
gives rise to a large value of the transmitted particle current and therefore
of the conductance.

For large values of $m_0$, we see that the conductance goes to zero for all
values of $\al$ and $\om$. This can be understood from Eq.~\eqref{expcond}.
Large $m_0$ means that $e^{-m_0}$ and therefore the lower component of
$\psi_{x=0+}$ is small. Hence the current, which is proportional to
$\psi_{x=0+}^\da \si^y \psi_{x=0+}$, will be small.

\subsubsection{Differential conductance as a function of $\al$}
\label{sec2b3}

\begin{figure}[H]
\centering
\includegraphics[scale=0.50]{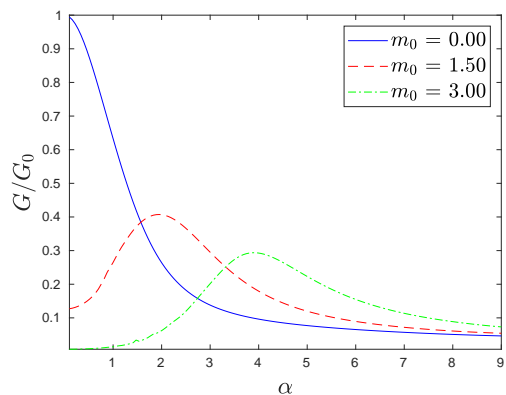}
\caption{$G/G_0$ as a function of $\al$ for different values of $m_0$, with
$E_0 = 2$ and $\om = 1.1$.} \label{fig04} \end{figure}

Figure~\ref{fig04} shows the dimensionless differential conductance $G/G_0$ as
a function of $\al$, for different values of $m_0$ and $\om$. Once again
we see that there are peaks in the conductance when $\al$ is close to $m_0$;
this can be understood in the same way as the peaks in Fig.~\ref{fig03}.
For large values of $\al$, we see in Fig.~\ref{fig05} that the
conductance approaches a constant value which is almost independent
of $m_0$. This can be understood since the magnetic barrier strength
is $m_0 + \al \cos (\om t)$; if $\al \gg m_0$, the barrier strength
is dominated by the $\al$ term. We note that for both small
and intermediate $\omega$ and small $m_0$, $G$ can be substantially
reduced by increasing the intensity of the applied radiation $\alpha$.
This provides optical control over the conductance; we will analyze
this point in more detail in the next section.

\subsubsection{Small $\om$ limit}
\label{sec2b4}

\begin{figure}[H]
\centering
\includegraphics[scale=0.6]{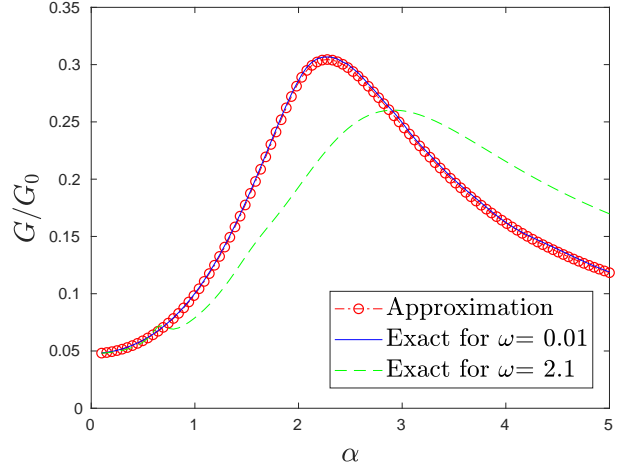}
\caption{$G/G_0$ as a function of $\al$. The current obtained by solving the
time-dependent problem is shown by the solid blue curve (for $\om = 0.01$) and
dashed green curve (for $\om =2.1$), while the value obtained by averaging
over $N=100$ equispaced values of $\phi$ is shown by the red circles. The
averaged results (red) agree well with the results for $\om=0.01$ (blue) which
is in the adiabatic regime, but do not agree with the results for $\om = 2.1$
(green).} \label{fig05} \end{figure}

Next, we look at the adiabatic limit $\hbar\om \ll E_0$. Since the barrier
strength varies very slowly with time in this limit, we can estimate the total
transmitted particle current value by studying the Hamiltonian at
an equispaced sequence of frozen times covering one time period $T$, and then
averaging over the results for all these times. Thus by solving a sequence
of problems with no time dependence, we expect to obtain a good approximation
of the time-dependent problem. The time-dependent term in the barrier strength
is $\al \cos(\om t)$. Denoting $\phi = \om t$, we fix $\phi$ at $N$
equispaced values, $\phi_j$, in the range 0 to $2\pi$, and find the total
transmitted current $I(\phi_j)$ for each value of $j$. We then calculate the
averaged current
\beq I_{avg} ~=~ \dfrac{\sum\limits_{j=1}^N I(\phi_j)}{N}, \eeq
and check how well this compares with the result for the time-dependent
problem. The results are shown in Fig.~\ref{fig05}. As we can see from the 
figure, the averaging approximation works well for $\om = 0.01$ which is much 
smaller than $E_0/\hbar$.
When $\om$ is larger than $E_0/\hbar$, we observe a significant difference
between from the averaged current and the actual current obtained for the
time-dependent system.

\subsubsection{Large $\om$ limit}
\label{sec2b5}

\begin{figure}[H]
\centering
\includegraphics[scale=0.60]{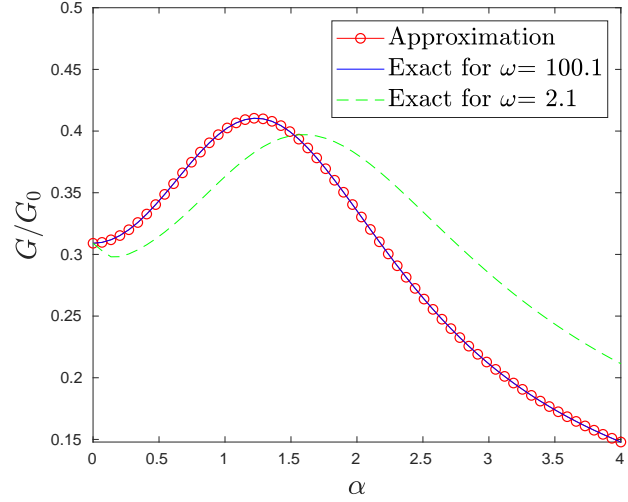}
\caption{$G/G_0$ as a function of $\al$. The solid blue and dashed green curves
show the currents obtained by solving the time-dependent problem for $\om$
equal to $100.1$ and $2.1$ respectively. The red circles show the approximate
solution obtained by using the wave functions in Eq.~\eqref{ef2}. We have
taken $E_0 = 2$ and $m_0=1$.} \label{fig06} \end{figure}

We now look at the opposite limit where $\hbar \om \gg E_0$. Since the energy
in the $n$-th band is given by $E_n = E_0 + n \hbar \om$, the gap between
successive side bands is much larger that $E_0$ in this limit. Hence we do not
expect the side bands to contribute much to the current. Also, the
momenta $k_{x,n}$ and $k_y$ are related to $E_n$ as $k_{x,n} = \pm (1/v)
\sqrt{E_n^2 - v^2 k_y^2}$; we can ignore $k_y$ in this expression if $E_n$ is 
very large. Thus $k_{x,n} \approx \pm E_n/v$. Eq.~\eqref{ef} then implies that 
the eigenfunctions of the side bands become independent of $n$, and we get
\beq \psi_n= \frac{1}{\sqrt{2}}\begin{pmatrix}
1 \vspace{0.5cm}\\ \dfrac{E_n}{ivk_{x,n}} \end{pmatrix}, \eeq
for $n \ne 0$, where we have ignored $k_y$. Simplifying this, we obtain
\bea \psi_{n,t}= \frac{1}{\sqrt{2}}\begin{pmatrix}
1 \vspace{0.5cm}\\ -i \end{pmatrix}
~~~\text{and}~~~
&\psi_{n,r}=\frac{1}{\sqrt{2}}\begin{pmatrix}
1 \vspace{0.5cm}\\ i
\end{pmatrix} \label{ef2} \eea
as the approximate wave functions in the large $\om$ limit for the transmitted
and reflected waves respectively for $n \ne 0$. Using these wave functions,
we solve the equations analogous to Eq.~\eqref{eqn1} to find $r_n$ and $t_n$
and hence the total transmitted particle current.
(Note that since the wave functions in Eq.~\eqref{ef2} do not depend
on $\om$, the transmitted current and conductance become independent of
$\om$ in the large $\om$ limit). Figure~\ref{fig06} shows a comparison
between the results obtained by this approximation and the exact result. We
see that the agreement is very good when $\om =100.1 \gg E_0/\hbar$ but
deviates significantly when $\om = 2.1$ is comparable to $E_0/\hbar$.

\section{Finite-width magnetic barrier}
\label{sec3}

In this section, we study the effects of a time-dependent magnetic barrier
which has a finite width; we will assume that the barrier lies in the region
$0 \le x \le L$. The incident and reflected waves lie in region $I$ where
$x < 0$, while the transmitted wave lies in region $III$ where $x > 0$.
In region $II$ where $0 \le x \le L$, the wave function satisfies the equation
\beq i\hbar \dfrac{\pa \psi}{\pa t} ~=~ \hbar v ~[ k_y \si^x ~+~ i \si^y
\frac{\pa}{\pa x} ~+~ \{ m_0+ V_0 \cos (\om t)\} ~\si^x]~ \psi.
\label{psi2} \eeq
We note that $m_0 = g \mu_B B_0/(\hbar v)$ and $V_0 = eA_0/(\hbar v)$ have the
dimensions of inverse length. We now assume that the solution of
Eq.~\eqref{psi2} is of the form
\beq \psi ~=~ \sum_{n=-\infty}^{\infty} ~\begin{pmatrix}
\al_n \\
\beta_n \end{pmatrix} e^{i(k'_x x ~+~ k'_y y ~-~ E_n t/\hbar)}, \eeq
where $E_n = E_0 +n \hbar \om$, $k'_y= k_y + m_0$, and $k'_x$ will be
determined as described below. Equating coefficients of
$e^{-i n \om t}$ on the two sides of Eq.~\eqref{psi2}, we obtain the following
equations
\bea -i\hbar vk'_y \al_n ~-~ \dfrac{i\hbar v V_0}{2}(\al_{n+1} ~+~ \al_{n-1})
~+~ iE_n \beta_n &=& \hbar vk'_x \al_n \non \\
~i\hbar vk'_y \beta_n ~+~ \dfrac{i\hbar v V_0}{2}(\beta_{n+1}+\beta_{n-1}) ~-~
iE_n\al_n &=& \hbar vk'_x \beta_n. \label{eqn2} \eea
If we truncate the above equations by keeping only $N$ bands, then
Eqs.~\eqref{eqn2} give an eigenvalue equation for $k'_x$ with $2N$ possible
eigenvalues; the eigenvalue equation looks as follows.
\beq \begin{pmatrix}
\ddots & \vdots & \vdots & \vdots & \vdots & \vdots & \vdots & \vdots &
\reflectbox{$\ddots$} \\
\cdots & -i\hbar vk'_y & iE_{-1} & -\frac{i\hbar v V_0}{2} & 0 & 0 & 0 &
\cdots \\
\cdots & -iE_{-1} & i\hbar vk'_y & 0 & \frac{i\hbar v V_0}{2} & 0 & 0 &
\cdots \\
\cdots & -\frac{i\hbar v V_0}{2} & 0 & -i\hbar vk'_y & iE_0 & -
\frac{i\hbar v V_0}{2} & 0 & \cdots \\
\cdots & 0 & \frac{i\hbar v V_0}{2} & -i E_0 & i\hbar vk'_y & 0 & \frac{i\hbar
v V_0}{2} & \cdots \\
\cdots & 0 & 0 & -\frac{i\hbar v V_0}{2} & 0 & -i\hbar vk'_y & iE_1 & \cdots \\
\cdots & 0 & 0 & 0 & \frac{i\hbar v V_0}{2} & -iE_1 & i\hbar vk'_y & \cdots \\
\reflectbox{$\ddots$} & \vdots & \vdots & \vdots & \vdots & \vdots & \vdots &
\vdots & \ddots \\
\end{pmatrix}
\begin{pmatrix}
\vdots \\ \al_{-1} \\ \beta_{-1} \\ \al_0 \\ \beta_0 \\ \al_1 \\ \beta_1 \\
\vdots \end{pmatrix} ~=~ \hbar v k'_x ~\begin{pmatrix}
\vdots \\ \al_{-1} \\ \beta_{-1} \\ \al_0 \\ \beta_0 \\ \al_1 \\ \beta_1 \\
\vdots \end{pmatrix}. \label{kx1} \eeq
After finding the different eigenvalues $k'_x$, denoted by $k'_{x,j}$, and
the corresponding values of $\al_n$ and $\beta_n$ denoted by $\al_{n,j}$ and
$\beta_{n,j}$, we proceed to find the reflection and transmission amplitudes
by matching wave functions at $x=0$ and $L$. At $x=0$, we have
\beq \begin{pmatrix}
1 \\ e^{-i\ta}
\end{pmatrix} \de_{n,0} ~+~ r_n \begin{pmatrix}
1 \\ \beta_n^- \end{pmatrix}
= \sum_{j=1}^{2N} ~C_j ~\begin{pmatrix}
\al_{n,j} \\
\beta_{n,j} \end{pmatrix}, \eeq
while at $x=L$, we get
\beq \sum_{j=1}^{2N} ~C_j ~\begin{pmatrix}
\al_{n,j} \\
\beta_{n,j} \end{pmatrix} e^{ik'_{x,j} L} ~=~ t_n \begin{pmatrix}
1 \\ \beta_n^+ \end{pmatrix} e^{ik_{x,n} L}, \eeq
where $k_{x,n}$ and $\beta_n^{\pm}$ are
given in Eqs.~\eqref{ef} and Eq.~\eqref{betan} respectively,
$C_j$'s are the coefficients of the wave
functions corresponding to different values of $k'_{x,j}$ in region $II$, and
$n$ takes $N$ possible values. The above equations give us a total of $4N$
conditions, since we have $N$ equations at both $x=0$ and $L$, and each wave
function has two components. We have to use these conditions to determine $4N$
quantities, namely, the $2N$ values of $C_j$ and $N$ values of both $r_n$ and 
$t_n$. We can therefore calculate all these quantities and thus determine the 
transmitted particle current in the region $x > L$ using Eq.~\eqref{it}.

\subsection{Numerical results}
\label{sec3a}

We now present our numerical results for the transmitted particle current and
conductance. In our numerical calculations, we will take the incident energy 
$E_0 = 2$ in units of $0.01$ eV, and the barrier width $L=1$ in units of 
$\hbar v / (0.02$ eV) $\simeq 17$ nm. The values of $m_0$ and $V_0$ will be 
given in units of $0.01$ eV/$(\hbar v) \simeq 0.03$ nm$^{-1}$.

\subsubsection{Transmitted particle current as a function of $\ta$}
\label{sec3a1}

\begin{figure}[H]
\centering
\subfigure[]{\includegraphics[scale=0.50]{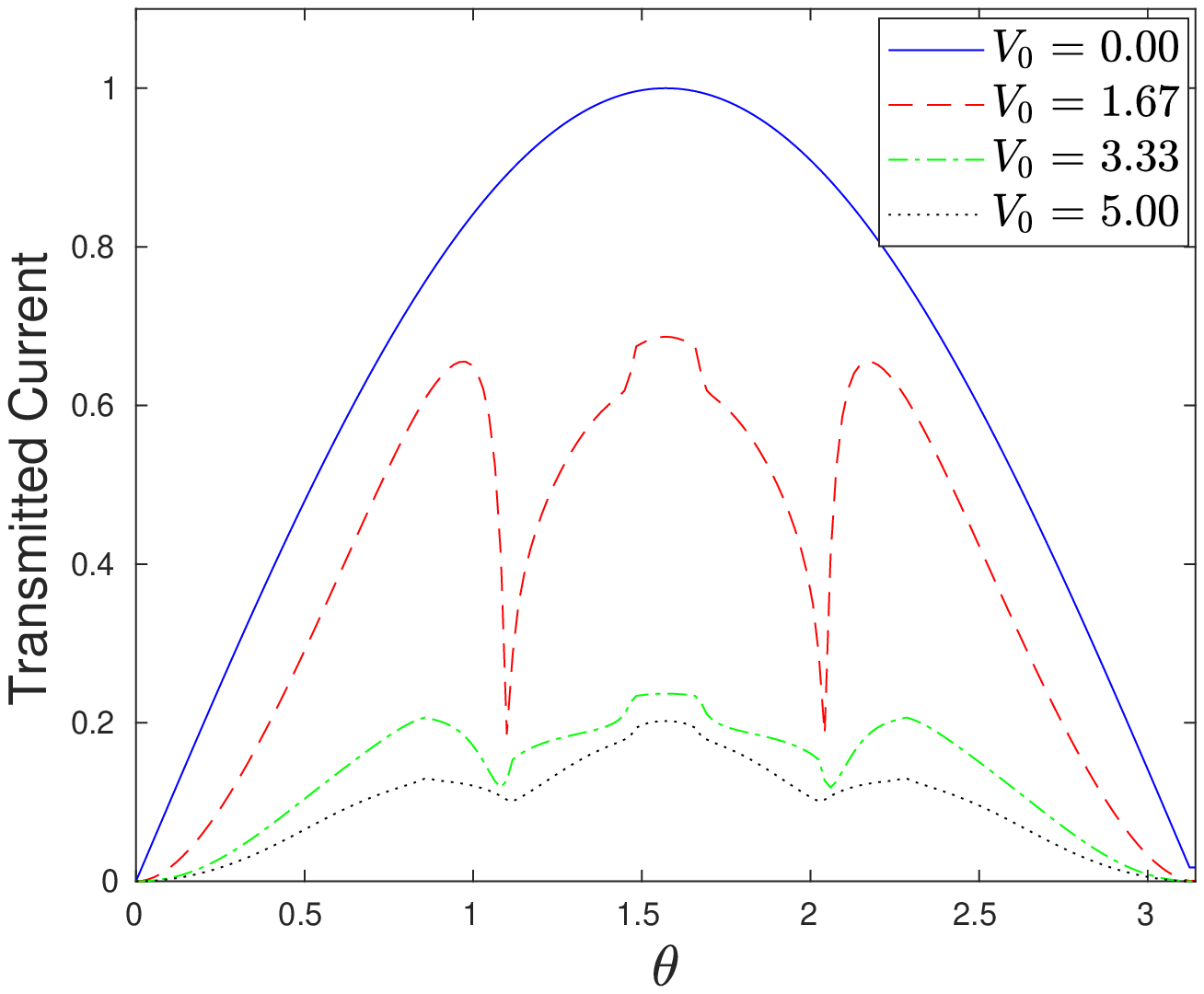}}
\subfigure[]{\includegraphics[scale=0.50]{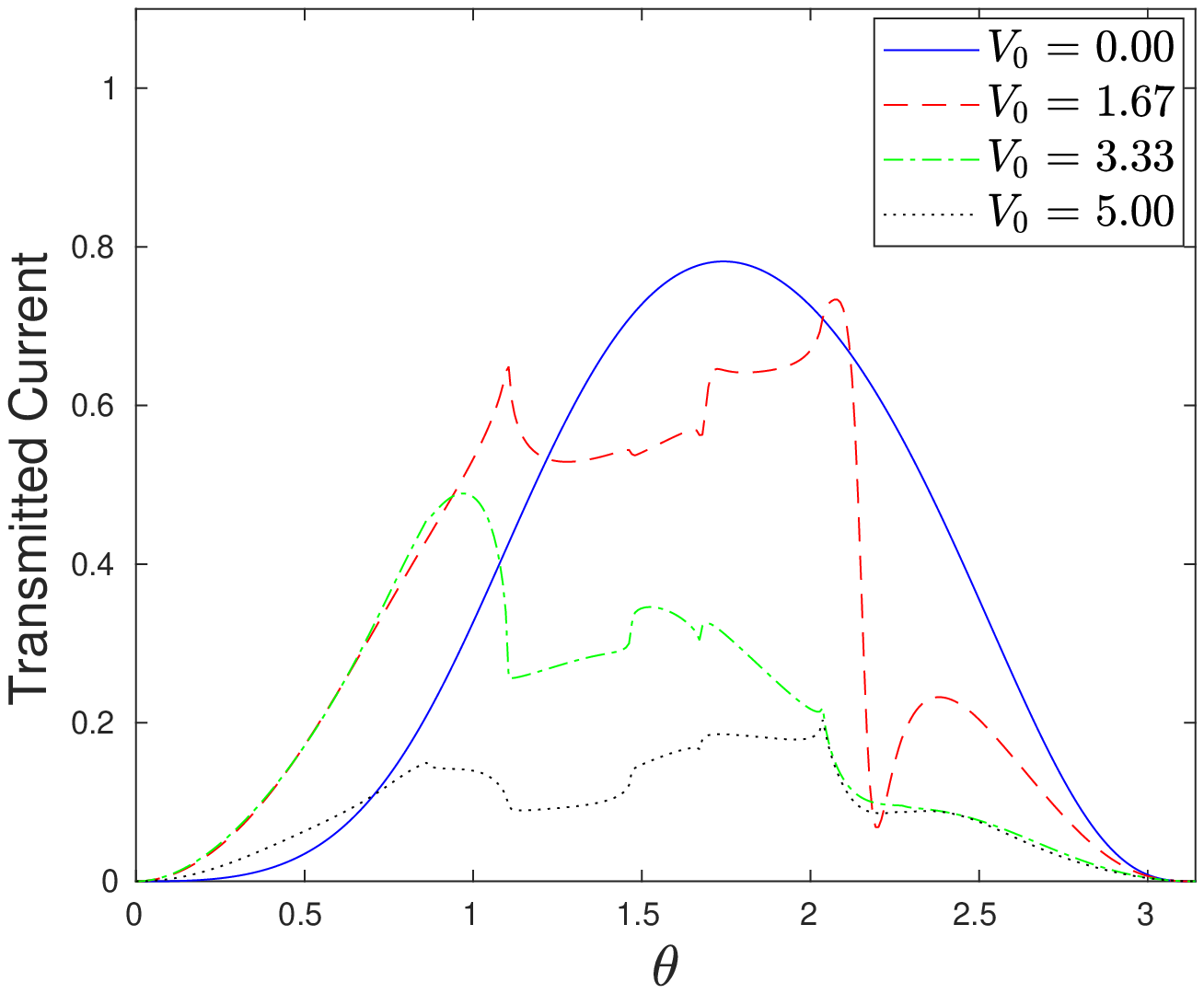}}
\caption{Transmitted particle current as a function of angle of incidence for
different values of $V_0$, with $E_0 = 2$, $L=1$, $\om= 1.1$ and $m_0$ equal
to (a) 0 and (b) 1.} \label{fig07} \end{figure}

We recall that $\ta = \tan^{-1}(k_x / k_y)$ is the angle of incidence in region
$I$. Figure~\ref{fig07} shows the transmitted particle current as function of
$\ta$, for different values of $V_0$ and $\om$, with $E_0 = 2$ and $m_0 = 0$
and 1. If $\om$ is not too large, we see kinks for certain values of $\ta$ for
the same reasons as discussed for the $\de$-function barrier. The plots are
symmetric about $\ta = \pi/2$ for $m_0 = 0$ (Fig.~\ref{fig07} (a)), but there
is no symmetry when $m_0 \ne 0$ (Fig.~\ref{fig07} (b)).


\subsubsection{Transmitted particle current as a function of $m_0$ and $V_0$}
\label{sec3a2}

\begin{figure}[H]
\centering
\subfigure[]{\includegraphics[scale=0.24]{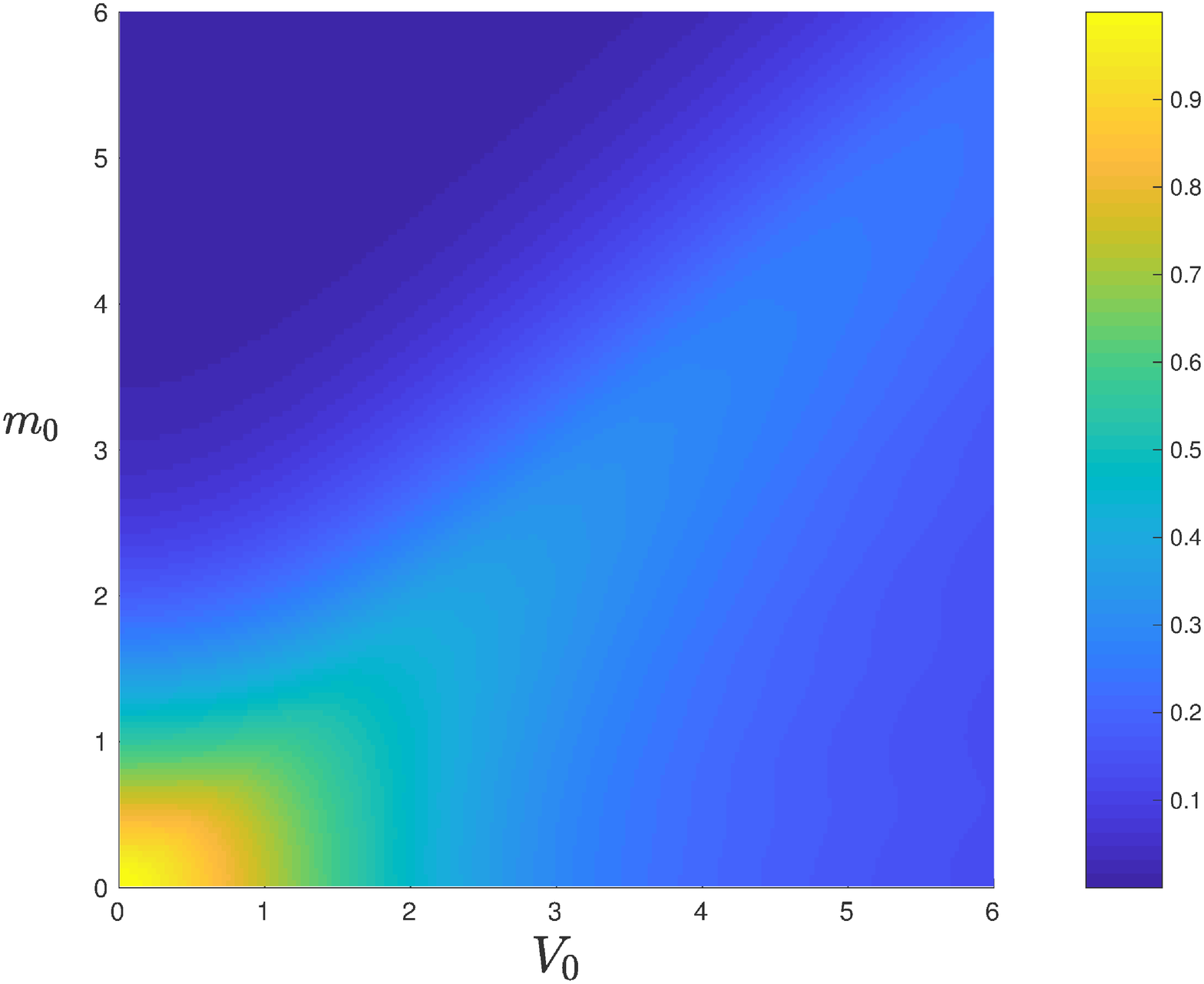}} 
\hspace*{.8cm}
\subfigure[]{\includegraphics[scale=0.24]{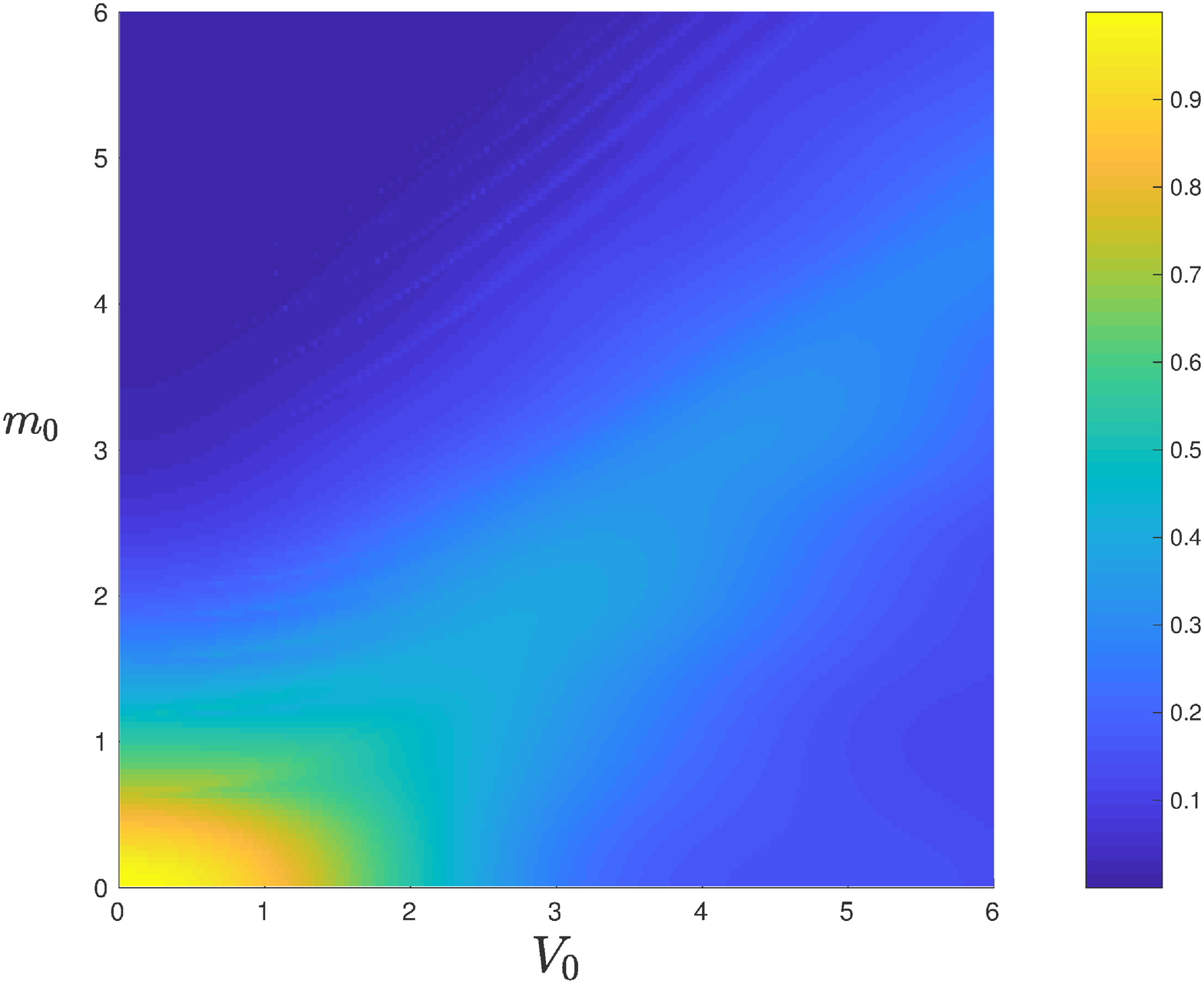}} 
\subfigure[]{\includegraphics[scale=0.24]{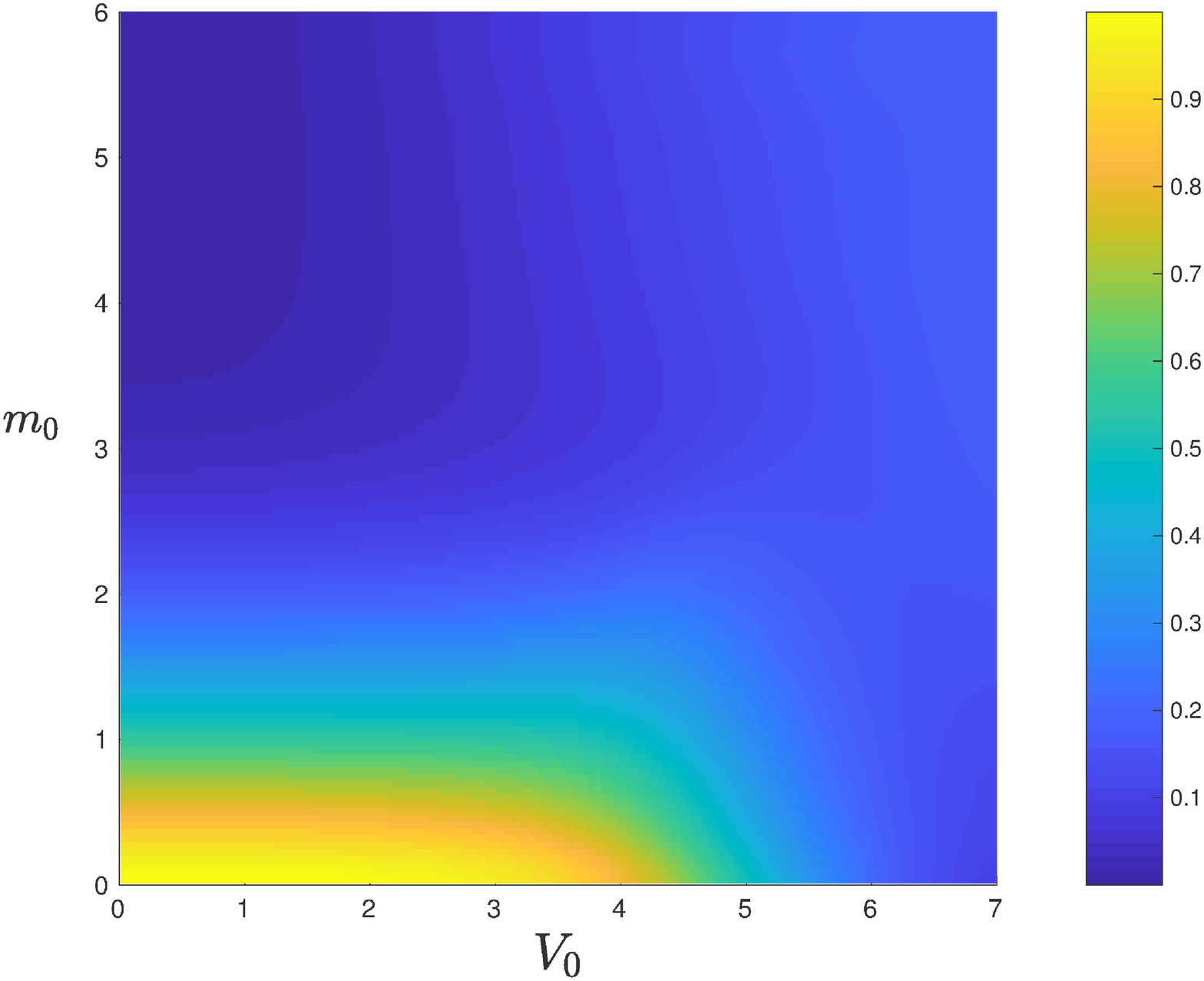}} 
\hspace*{.8cm}
\subfigure[]{\includegraphics[scale=0.24]{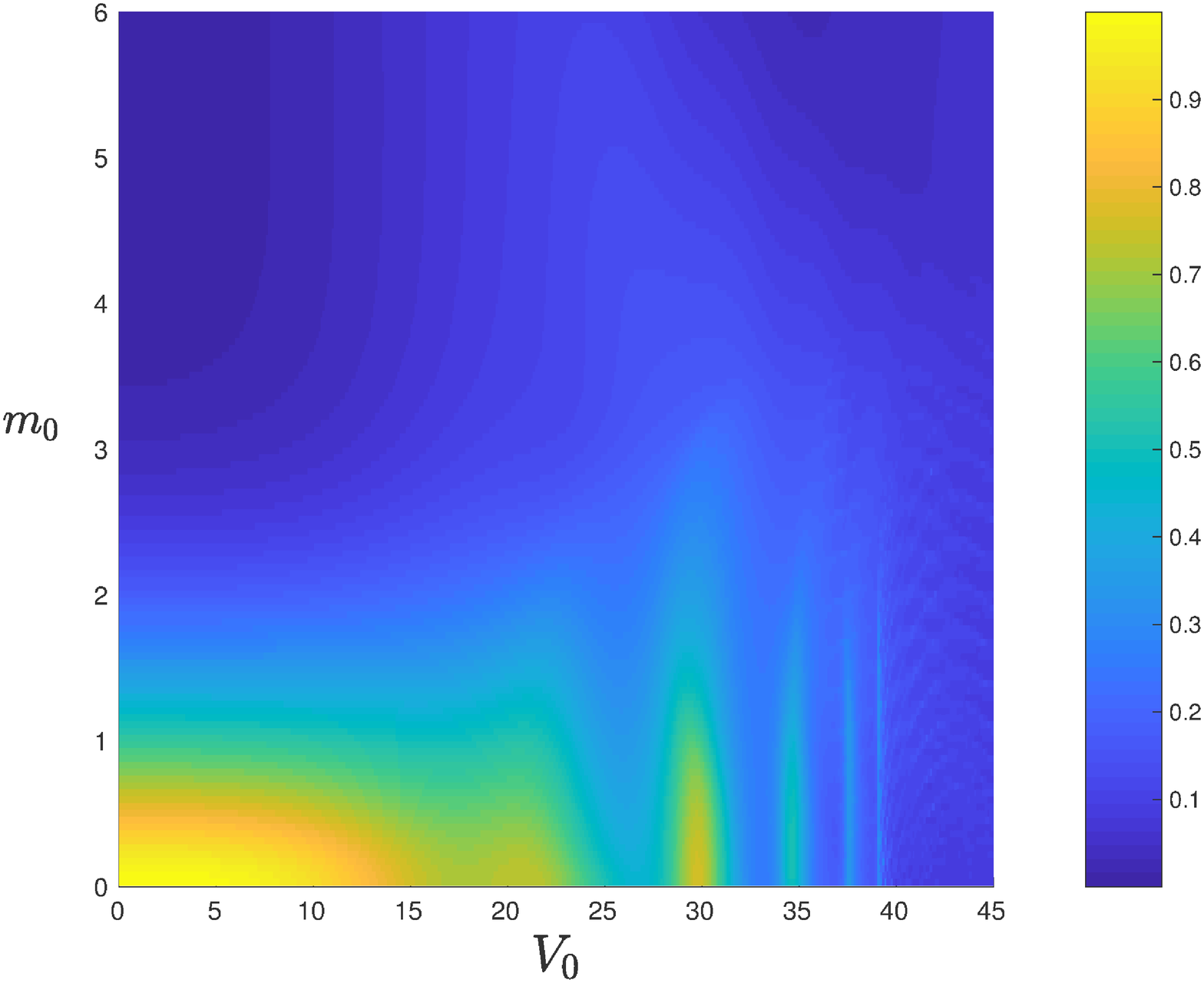}} 
\caption{Surface plots of $G/G_0$ as a function of $V_0$ and $m_0$,
for $E_0 = 2$, $L=1$, and $\om$ equal to (a) $0.01$, (b) $1.1$, (c) $10.1$, and
(d) $40.1$.} \label{fig08} \end{figure}

It is interesting to look at surface plots of the dimensionless differential
conductance $G/G_0$ for different values of $\om$, with $E_0 = 2$ and $L=1$.
These are shown in Fig.~\ref{fig08} for $\om = 0.01, ~1.1, ~10.1$ and $40.1$.

For the smallest value of $\om = 0.01$, we see that the current is
maximum around the line $m_0= V_0$. This can be understood by an
argument similar to the one used to understand the positions of the
peaks in Fig.~\ref{fig03}. If $\om$ is small, the strength of the
magnetic barrier, given by $m_0 + V_0 \cos (\om t)$ stays close to
its extreme values of $m_0 \pm V_0$ for a long time. For instance,
if $m_0 = V_0$, an expansion of $t$ around the time $t_0 = \pi/\om$
gives $\hbar v [m_0 + V_0 \cos (\om t)] ~\simeq~ \hbar v m_0 w^2 (t
- t_0)^2/2$, which is much smaller than the incident energy $E_0$
for a duration of time $|t - t_0|$ which is of the order of $(1/\om)
\sqrt{E_0/ (\hbar v m_0)}$. If this time is much larger than the
time taken by the electron to go across the barrier region, namely,
$L/v$, we expect the conductance would be large since the electron
sees a very small barrier strength.

For large values of $\om$ as shown in Fig.~\ref{fig08} (c), we observe that a
new phenomenon emerges. Namely, we find that close to the line $m_0 = 0$,
the conductance oscillates significantly with $V_0$. In particular, there
are peaks in $G/G_0$ for certain values of $V_0$ which are reminiscent of
resonances in transmission through a barrier. To obtain a better
understanding of these peaks, we will make some simplifying assumptions
about region $II$. Since the peaks appear even for very small values of
$m_0$ and for large values of $V_0 \gg E_0$, we will set $m_0=0$
and analyze the problem both numerically and using perturbation theory.

\begin{figure}[H]
\centering
\subfigure[]{\includegraphics[scale=0.25]{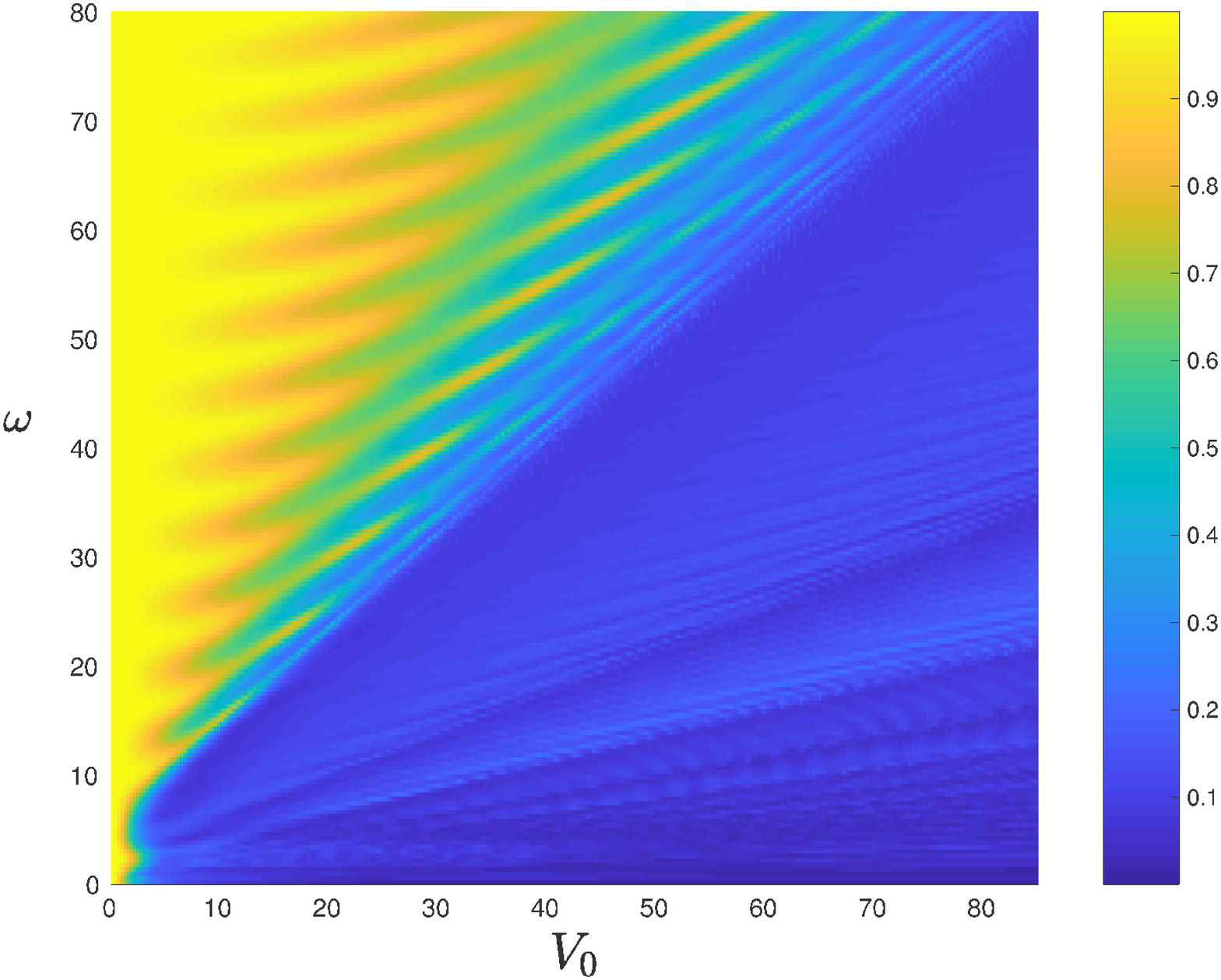}}
\hspace*{.8cm}
\subfigure[]{\includegraphics[scale=0.25]{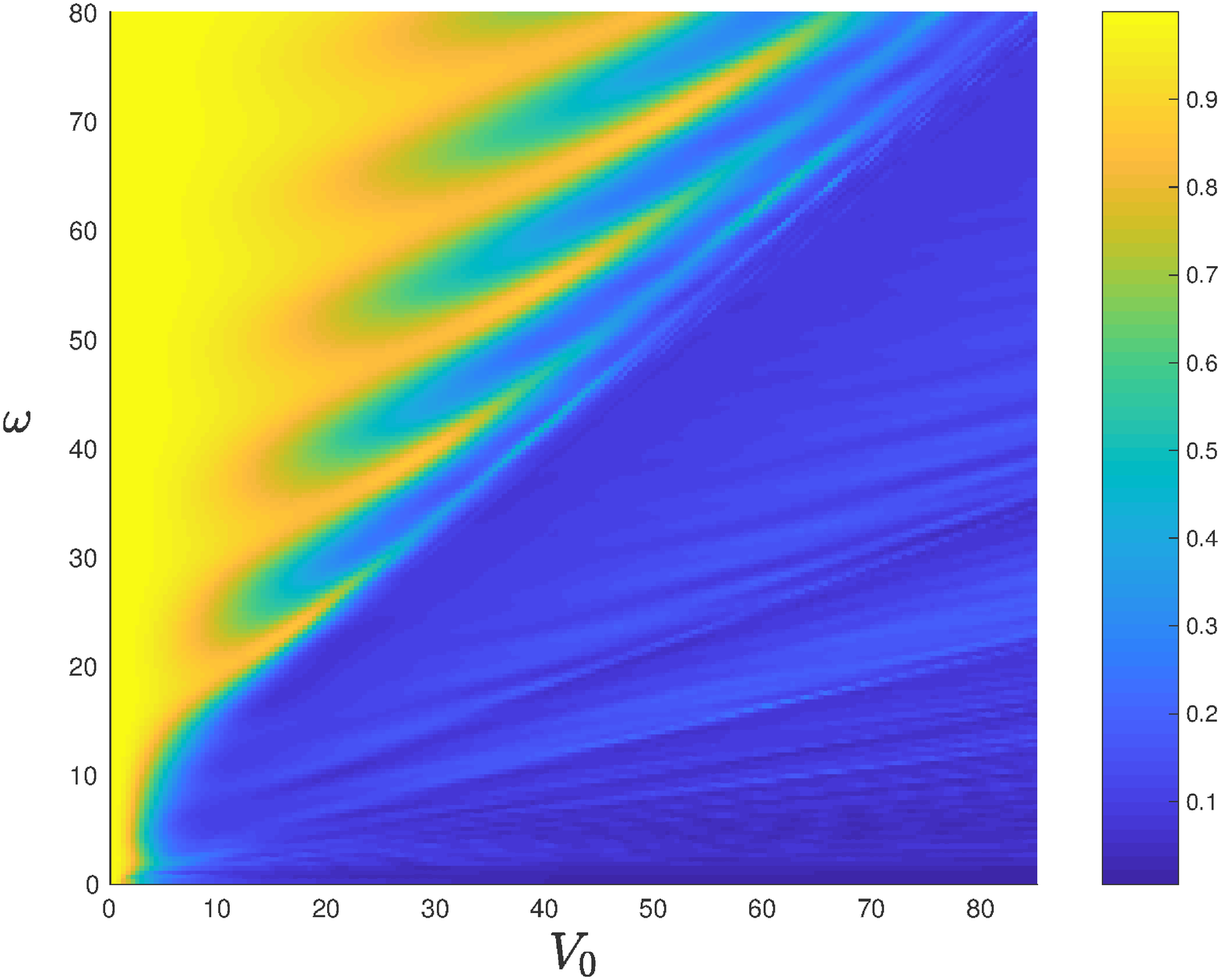}}
\caption{Surface plots of $G/G_0$ as a function of $V_0$ and $\om$, for $E_0
= 2$, $m_0 = 0$, and (a) $L=1$ and (b) $L=0.5$.} \label{fig09} \end{figure}

Figure~\ref{fig09} shows a surface plot of $G/G_0$ as a function of
$V_0$ and $\om$, for $m_0 = 0$, $E_0 = 2$, and (a) $L=1$ and (b)
$L=0.5$. In the upper left parts of the figures, we see prominent
oscillations in the conductance, while in the lower right parts, we
see that the conductance is small everywhere but there are straight
lines along which the conductance is particularly small. We will
provide an analytical understanding of both these features below.

\begin{figure}[H]
\centering
\includegraphics[scale=0.5]{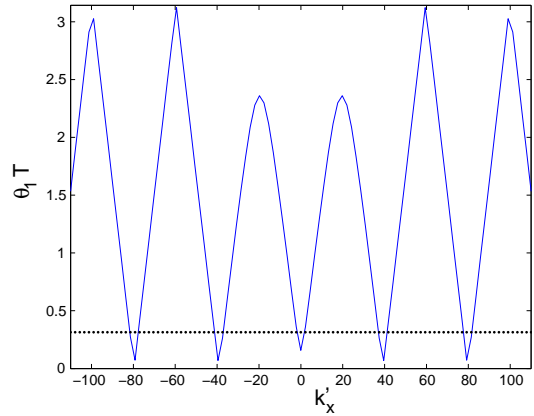}
\caption{Plot of $\ta_1 T$ as a function of $k'_x$ obtained
from a numerical evaluation of the Floquet operator $U$, for $E_0 = 2$, $k_y
=1$, $\om = 40$, $V_0 = 10$, and $m_0 = 0$. The horizontal dotted line lies
at $E_0 T /\hbar \simeq 0.314$.} \label{fig10} \end{figure}

Inside the barrier (region $II$), the Hamiltonian is given by Eq.~\eqref{psi2},
\beq H(t) ~=~ \hbar v ~[ k_y ~\si^x ~-~ k'_x ~\si^y ~+~ V_0 \cos (\om t) ~
\si^x], \label{hamt2} \eeq
where $k'_x$ denotes one of the possible values of the momentum of
an electron inside the barrier.
We can numerically find the Floquet operator $U$ defined in Eq.~\eqref{flo1}
and find its eigenvalues which have the form $e^{\pm i \ta_1 T}$, where
we take $\ta_1 T$ to lie in the range $[0,\pi]$. A plot of $\ta_1 T$ versus
$k'_x$ is shown in Fig.~\ref{fig10} for $E_0 = 2$, $k_y =1$, $\om = 40$,
$V_0 = 10$, and $m_0 = 0$. The horizontal dotted line lies at the value
$E_0 T/\hbar \simeq 0.314$. Since the Floquet eigenvalue inside the barrier
must match the eigenvalue outside (regions $I$ and $III$), the intersections
of the horizontal line with the plot of $\ta_1 T$ shows the values that
$k'_x$ can take. We note from the figure that the three smallest possible 
values of $k'_x$ lie near zero and $\pm \om/v$. We will now derive these 
values analytically using Floquet perturbation theory~\cite{soori}.

To develop the perturbation theory, we write the Hamiltonian in
Eq.~\eqref{hamt2} as
\beq H(t) ~=~ H_0 (t) ~+~ V (t), \eeq
where $H_0$ is much larger than $V$. We will assume that $H_0 (t)$ commutes 
with itself at different times; hence its eigenstates are time-dependent 
although its eigenvalues may vary with time. In our problem described by 
Eq.~\eqref{psi2}, we will consider two cases given by
\bea H_0 &=& \hbar v V_0 \cos (\om t) ~\si^x, \non \\
V &=& \hbar v ~(k_y ~\si^x ~-~ k'_x ~\si^y), \label{h0v1} \eea
where we assume $V_0 \gg k_y, ~k'_x$, and
\bea H_0 &=& - \hbar v k'_x ~\si^y, \non \\
V &=& \hbar v ~[k_y ~+~ V_0 \cos (\om t)] ~\si^x, \label{h0v2} \eea
where we assume $k'_x \gg k_y, ~V_0$. We now consider the two cases in turn.

For the case given in Eq.~\eqref{h0v1}, the unperturbed problem given by
$i \hbar \pa \psi /\pa t = H_0 \psi$ has solutions of the form
\bea \psi_1 (t) &=& \phi_1 ~\exp [-i (v V_0/\om) ~\sin (\om t)], \non \\
\psi_2 (t) &=& \phi_2 ~\exp [i (v V_0 /\om) ~\sin (\om t)], \non \\
{\rm where} ~~~\phi_1 &=& \frac{1}{\sqrt{2}} ~\begin{pmatrix}
1 \\ 1 \end{pmatrix} ~~~{\rm and}~~~ \phi_2 ~=~ \frac{1}{\sqrt{2}} ~
\begin{pmatrix} 1 \\ -1 \end{pmatrix}, \eea
and the corresponding eigenvalues of $H_0 (t)$ are given by $E_1(t)= \hbar v
V_0 \cos (\om t)$ and $E_2(t)= - \hbar v V_0 \cos (\om t)$ respectively.
Since these satisfy the condition
\beq e^{(i/\hbar) \int_0^T dt ~[E_1(t)- E_2(t)]} ~=~ 1, \label{reso} \eeq
we have to use degenerate perturbation theory~\cite{soori}.
Assuming that the wave function has the form
\beq \psi(t)= \sum_{p=1}^2 C_m(t) e^{-(i/\hbar) \int_0^t dt' E_m (t')}
\phi_m, \label{cmt} \eeq
and solving for the equation $i \hbar \pa \psi /\pa t = H(t) \psi$,
we obtain a differential equation for $C_m (t)$. Integrating from $t=0$
to $T$, we have, to first order in $V$,
\beq C(T) ~=~ (I ~-~ iM) ~C(0), \eeq
where $C(t) = (C_1 (t), C_2 (t))$ is a column, and the matrix $M$ has elements
\beq M_{mn} ~=~ \frac{1}{\hbar} ~\int_0^T dt ~\psi_m^\dag V \psi_n ~
e^{i\int_0^t ~dt' (E_m (t')-E_n(t'))}. \label{mmn} \eeq
In our problem, $M$ is a $2 \times 2$ matrix with elements
\bea M_{11} &=& v k_y T, ~~~M_{22} ~=~ - v k_y T, \non \\
M_{12} &=& i J_0\big({\frac{2V_0}{\om}}\big) ~v k'_x T, ~~~ M_{21} ~=~ - ~i
J_0\big(\frac{2V_0}{\om}\big) ~ v k'_x T. \eea
Up to first order in $V$, the eigenvalues of $M$ are given by
\beq \lm_\pm ~=~ \pm ~v T \sqrt{k_y^2 ~+~ [J_0(2V_0 /\om) k'_x]^2}. \eeq
Hence we can find eigenstates of $M$ such that
\beq C_\pm (T) ~=~ e^{i\lm_\pm} ~C_\pm (0). \eeq
Since the Hamiltonian is periodic, Floquet theory implies
that \beq \psi_n (T) ~=~ e^{- i \ta_n T} \psi_n (0). \eeq
This gives a relation between $\ta_\pm$ and $\lm_\pm$; in our case it is
\beq \ta_\pm ~=~ \pm ~v \sqrt{k_y^2 ~+~[J_0(2V_0 /\om) k'_x]^2}. \eeq
Since the Floquet eigenvalues in regions $I$ and $III$ given by
$e^{-iE_0 T/\hbar}$ (where $E_0$ is the energy of the incident electron) must
be equal to the Floquet eigenvalue in region $II$ (the barrier), we obtain an
expression for two of the allowed values of $k'_x$, namely,
\beq k'_x ~=~ \pm ~\frac{\sqrt{(E_0/\hbar v)^2 ~-~ k_y^2}}{J_0(2V_0 /\om)}.
\label{kpx1} \eeq
This gives an expression for $k'_x$ of the order of $E_0/(\hbar v)$ which lies
near zero in Fig.~\ref{fig10} since we have taken $E_0 \ll \hbar \om$ in that
figure.

\begin{figure}[H]
\centering
\subfigure[]{\includegraphics[scale=0.25]{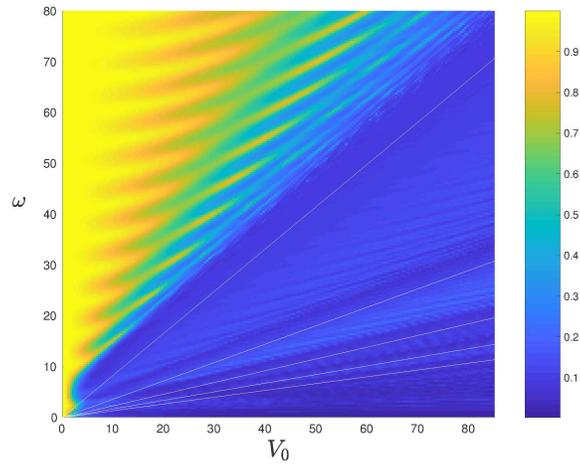}}
\hspace*{.8cm}
\subfigure[]{\includegraphics[scale=0.25]{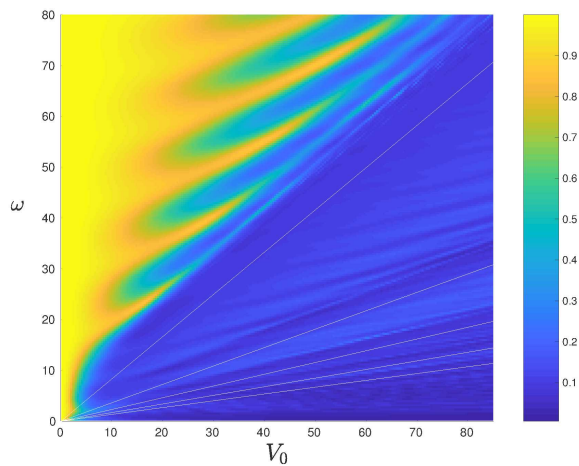}}
\caption{Surface plots of $G/G_0$ as a function of $V_0$ and $\om$, for $E_0
= 2$, $m_0 = 0$, and (a) $L=1$ and (b) $L=0.5$. These are the same as
Figs.~\ref{fig09} (a) and (b), but we have added some straight white lines
corresponding to $J_0 (2V_0/\om) = 0$; from top to bottom, these lines have
slopes given by $0.832, 0.362, 0.231, 0.170$ and $0.134$.}
\label{fig11} \end{figure}

Eq.~\eqref{kpx1} implies that if $J_0(2V_0 /\om) = 0$, there is no
value of $k'_x$ for which the Floquet eigenvalue in region $II$ will
match the Floquet eigenvalues in regions $I$ and $III$, except for
the special case $k_y = \pm E_0/(\hbar v)$ which corresponds to
grazing incidence where the transmission probability is always zero.
Hence, it is not possible to an electron to transmit through region
$II$ if the parameters $(V_0,\om)$ lie on one of the straight lines
where $J_0 (2V_0 /\om) = 0$. The first five zeros of $J_0 (z)$ are
given by $z = 2.405, 5.520, 8.654, 11.792$ and $14.931$, which give
$\om/V_0 = 0.832, 0.362, 0.231, 0.170$ and $0.134$. We see in
Fig.~\ref{fig11} that the conductance is indeed particularly small
on the white lines whose slopes correspond to zeros of $J_0 (2V_0 /\om)$.

We now turn to the case given in Eq.~\eqref{h0v2}. The unperturbed problem
given by $i \hbar \pa \psi /\pa t = H_0 \psi$ has solutions of the form
\bea \psi_1 (t) &=& \phi_1 ~e^{-i v k'_x t}, \non \\
\psi_2 (t) &=& \phi_2 ~e^{i v k'_x t}, \non \\
{\rm where} ~~~\phi_1 &=& \frac{1}{\sqrt{2}} ~\begin{pmatrix}
1 \\ -i \end{pmatrix} ~~~{\rm and}~~~ \phi_2 ~=~ \frac{1}{\sqrt{2}} ~
\begin{pmatrix} 1 \\ i \end{pmatrix}, \eea
and the corresponding eigenvalues of $H_0$ are given by $E_1 = k'_x$
and $E_2 = - k'_x$ respectively. These do {\it not} satisfy the condition
in Eq.~\eqref{reso} if $k'_x \ne n \om /(2v)$ for any integer value of $n$.
We can then use non-degenerate perturbation theory~\cite{soori}.
Assuming that the wave function has the form
\beq \psi(t)= \sum_{p=1}^2 C_m(t) e^{-(i/\hbar) \int_0^t dt' E_m (t')}
\phi_m, \eeq
and solving for the equation $i \hbar \pa \psi /\pa t = H(t) \psi$,
we obtain the following coupled differential equations
\bea \frac{dC_1}{dt} &=& e^{i2 v k'_x t} ~v~[k_y ~+~ V_0 \cos (\om t)] ~C_2,
\non \\
\frac{dC_2}{dt} &=& -~ e^{-i2 v k'_x t} ~v~[k_y ~+~ V_0 \cos (\om t)] ~C_1.
\eea
To solve these equations, we have to choose the initial conditions
$(C_1 (0), C_2 (0))$. To find the change in the Floquet eigenvalue
$e^{-i \ta_1 T}$ of a Floquet eigenstate which lies close to $\psi_1 (t)$,
we choose $C_1 (0) = 1$ and $C_2 (0)$ to be small and of order $k_y, ~V_0$.
Demanding that $\psi (T) = e^{-i \ta_1 T} \psi (0)$, we discover that
$\ta_1$ differs from $k'_x T$ only at second order in the perturbation, namely,
\beq \ta_1 ~=~ v ~\left[ k'_x ~+~ \frac{k_y^2}{2 k'_x} ~+~ \left( \frac{V_0}{2}
\right)^2 ~\frac{1}{2k'_x ~+~ \om/v} ~+~ \left( \frac{V_0}{2} \right)^2~
\frac{1}{2k'_x ~-~ \om/v} \right], \label{ta1} \eeq
modulo integer multiples of $\om$ since $\ta_1$ is a periodic variable with
period $\om$.
A similar calculation for a Floquet eigenstate which lies close to $\psi_2 (t)$
shows that $\ta_2 = - \ta_1$. Eq.~\eqref{ta1} shows that the second order
correction is small except around the points $k'_x = 0$ and $\pm \om /(2v)$.

Next, we can study what happens if $k'_x = n \om /(2v)$ where the condition
in Eq.~\eqref{reso} is satisfied. We then have to use degenerate perturbation
theory as described in Eqs.~(\ref{cmt}-\ref{mmn}). We discover that the matrix 
$M$ defined in Eq.~\eqref{mmn} is identically equal to zero if $k'_x \ne 
0, ~\pm \om /(2v)$. We can therefore use non-degenerate perturbation theory 
and go up to second order as in Eq.~\eqref{ta1}. We thus conclude that
Eq.~\eqref{ta1} holds for any value of $k'_x$ except around 0 and $ \pm
\om /(2v)$.

We now study the region around $k'_x = \pm \om /v$ where we have pairs of
solutions for $\ta_1 T = E_0 T /\hbar$ as we see in Fig.~\ref{fig10}. We can
find an expression for the two solutions near $k'_x = \om /v$ by replacing
$k'_x$ by $\om/v$ in the last three terms in Eq.~\eqref{ta1} (which are of
second order in the perturbation) but not in the first term.
Eq.~\eqref{ta1} then implies that the two possible values of $k'_x$
which yield $\ta_1 = E_0 /\hbar$ modulo $\om$ are given by
\beq k'_x ~=~ \frac{\om}{v} ~\pm~ \frac{E_0}{\hbar v} ~-~
\frac{v k_y^2}{2 \om} ~-~ \frac{v V_0^2}{3 \om}. \label{kpx2} \eeq
Since $E_0 \ge \hbar v |k_y|$, and we are working in the regime $\hbar \om
\gg E_0$, we can ignore the term $vk_y^2/(2\om)$. We then get
\beq k'_x ~=~ \frac{\om}{v} ~\pm~ \frac{E_0}{\hbar v} ~-~
\frac{v V_0^2}{3 \om}. \label{kpx3} \eeq

\begin{figure}[H]
\centering
\includegraphics[scale=0.25]{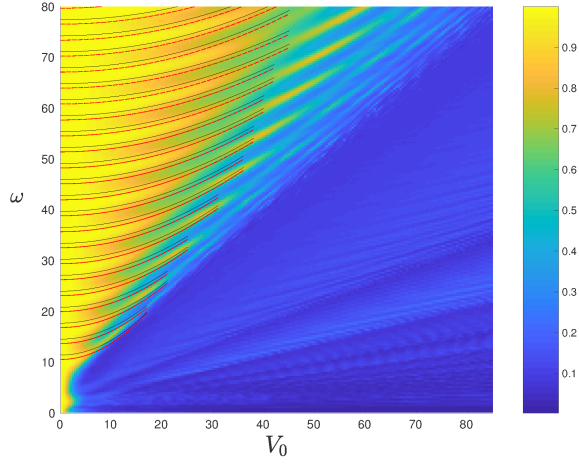} 
\caption{Surface plot of $G/G_0$ as a function of $V_0$ and $\om$, for $E_0
= 2$, $m_0 = 0$, and $L=1$. This is the same as Fig.~\ref{fig09} (a), but we
have added black and red lines which are given by Eq.~\eqref{om1}, where the
second term is $\pm E_0/\hbar$ respectively, and the integer $p$ increases
as we go up from the bottom to the top.} \label{fig12} \end{figure}

In problems involving transmission through a barrier, we typically
find that the condition for resonances is given by $e^{i k'_x L}=
\pm 1$, where $k'_x$ is the momentum inside the barrier (see, for
instance, Ref.~\onlinecite{schwabl}). This requires that $k'_x = p
\pi/L$, where $p = 1, 2, 3, \cdots$. Since the last two terms in
Eq.~\eqref{kpx3} are small, we obtain the approximate expression
\beq \om ~=~ \frac{\pi p v}{L} ~\pm~ \frac{E_0}{\hbar} ~+~ \frac{vL
V_0^2}{3 \pi p}. \label{om1} \eeq This implies that in a plot of the
conductance versus $(V_0,\om)$, the resonance regions (large
conductance) will have a spacing given by $\pi v /L$ when $V_0 = 0$
and will curve up as $V_0^2$ when $V_0$ is increased, provided that
$V_0/\om$ is small. This is exactly what we see in Fig.~\ref{fig12}
where the black and red lines are given by Eq.~\eqref{om1}, where
the second term is $\pm E_0 /\hbar$ respectively, and the integer $p$ 
increases as we go up from the bottom to the top. In the figure, we see 
that the spacing between either the black lines or the red lines is 
$\pi v/L = \pi$ at $V_0 = 0$. The black and red lines go up quadratically 
with increasing $V_0$ with the correct curvatures as given in Eq.~\eqref{om1}.

\subsubsection{Effective time-independent magnetic barrier}
\label{sec3a3}

In this section, we will map the time-dependent system with a given set of
parameter values $(m_0, ~V_0, ~\om)$ to a time-independent system with a
magnetic barrier which has only the term $m_0 \si^x$; we do the mapping
by demanding that the two systems should have the same value of the
differential conductance. The procedure is as follows. On the one hand,
we will calculate the conductance of the
time-dependent system and find its dependence on $m_0$ and $\om$, keeping
$V_0$ fixed. On the other hand, we will calculate the conductance for the
time-independent system and find its dependence on the single parameter $m_0$.
We will then use the condition of equal conductance to find an effective
value of $m_0$, called $m_{eff}$, of the time-independent system as a
function of $\om$ and $m_0$ of the time-dependent system.

\begin{figure}[H]
\centering
\includegraphics[scale=0.5]{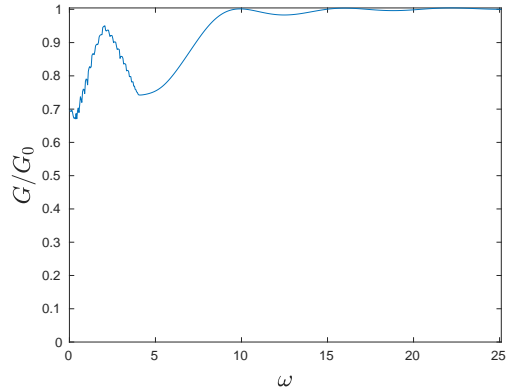}
\caption{$G/G_0$ as a function of $\om$ for $V_0= 1.1$, $E_0= 2$, $L=1$, and
$m_0=0$.} \label{fig13} \end{figure}

\begin{figure}[H]
\centering
\includegraphics[scale=0.5]{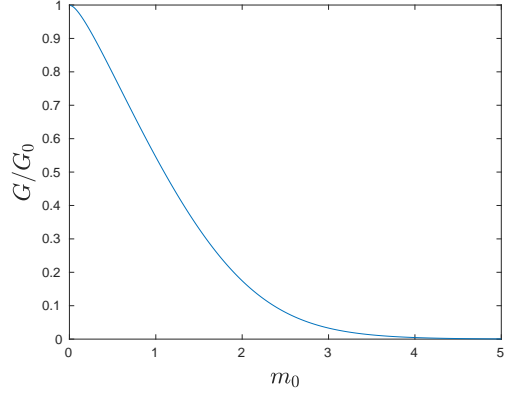}
\caption{$G/G_0$ as a function of $m_0$ in a time-independent system with
$E_0 =2$ and $L=1$.} \label{fig14} \end{figure}

We therefore first look at the time-dependent system and calculate the
conductance as a function of $\om$ for different
values of $m_0$ and fixed $V_0=1.1$. This is shown in Fig.~\ref{fig13}.
Next, we consider a time-independent system with a barrier
strength $m_0$. The Hamiltonian in region $II$ of this system is given by
\beq i \hbar \dfrac{\pa \psi}{\pa t} ~=~ \hbar v ~[k_y \si^x ~-~ k'_x \si^y
~+~ m_0 \si^x]~ \psi. \eeq
The conductance of this system as a function of $m_0$ is shown in
Fig.~\ref{fig14}.

\begin{figure}[H]
\centering
\includegraphics[scale=0.5]{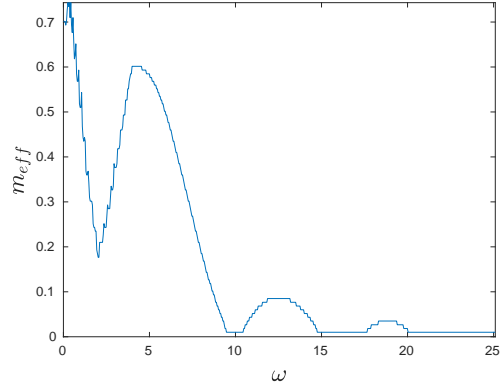}
\caption{$m_{eff}$ of the time-independent problem as a function of $\om$, for
$V_0= 1.1$, and $m_0=0$ in the time-dependent system. We have taken $E_0 = 2$
and $L=1$.} \label{fig15} \end{figure}

Finally, we can use Figs.~\ref{fig13} and \ref{fig14} to map the
time-dependent system with some values of $\om$ and $m_0$ to a
time-dependent system with a barrier strength $m_{eff}$, by
demanding that the two systems should have the same conductance.
Fig.~\ref{fig15} shows $m_{eff}$ as a function of $\om$, for $V_0 =
1.1$ and $m_0 = 0$ in the time-dependent system. We have set $E_0 =
2$ in all cases. In Fig.~\ref{fig15} we see that $m_{eff}$
approaches a constant as $\om$ becomes large. This can be understood
using the fact that when $m_0$ and $V_0$ are fixed, the conductance
of time-dependent system tends to a constant when $\om$ becomes much
larger than all the other energy scales in the problem. The reason
for this is similar to the one given in Sec.~\ref{sec2b5} for the
case of a $\de$-function magnetic barrier. We note from Fig.\ \ref{fig15}
that $m_{eff}$ is large for small frequencies whereas it goes to zero for 
large frequencies. This allows for a frequency induced control over the 
barrier conductance for a fixed amplitude. For large frequencies, one expects 
the junction to be conducting (since $m_{eff} \simeq 0$) while for small 
frequencies one has large $m_{eff}$ which will lead to $G \to 0$.

%
%
%

\subsubsection{Conductance as a function of $E_0$}
\label{sec3a5}

\begin{figure}[H]
\centering
\includegraphics[scale=0.60]{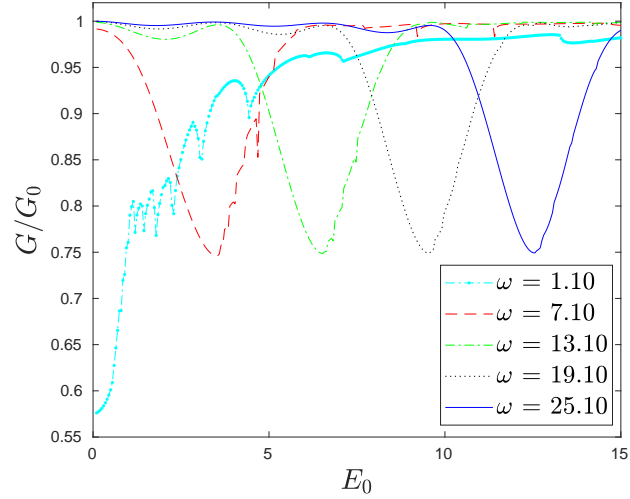}
\caption{$G/G_0$ as a function of $E_0$ for different values of barrier
strength $\om$, for $L=1$, $m_0= 0$ and $V_0=1.1$.} \label{fig16} \end{figure}

Finally, we study the conductance as a function of the incident energy
$E_0$ for different values of $\om$ and some fixed values of $m_0$ and $V_0$.
We find an interesting result that there is a dip in the conductance around
$E_0 = \hbar \om /2$; the dips are quite prominent when $\om$ is large.
This is shown in Fig.~\ref{fig16}.

\begin{figure}[H]
\centering
\includegraphics[scale=0.5]{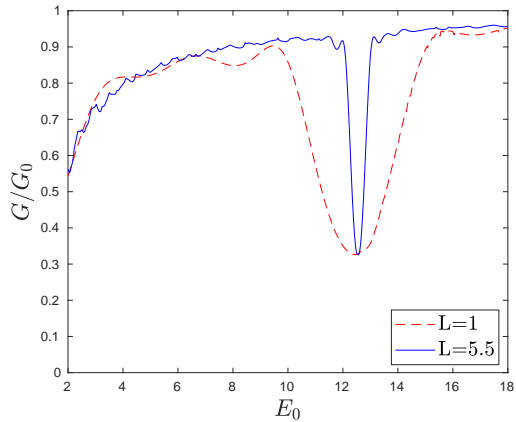}
\caption{$G/G_0$ as a function of $E_0$ for $m_0=1$, $\om = 25.1$, and $L= 1$
and $5.5$, keeping $V_0 L= 2.3$ fixed.} \label{fig17} \end{figure}

These dips can be understood as follows. When $E_0=\hbar \om /2$, we have
$E_{-1} = E_0 - \hbar \om = -E_0$, namely, the energies of the bands $n=0$
and $-1$ become
equal in magnitude. We then find numerically that the maximum contribution to
the conductance comes from only these two bands. Hence ignoring the
contributions from all the other bands, we can proceed to study this problem
analytically. To find the allowed values of $k'_x$ inside the barrier, we have
to solve the eigenvalue equation as in Eq.~\eqref{kx1}, but now with just
two bands. The equation then takes the form
\beq \begin{pmatrix} -i\hbar vk_y & iE_{-1} & -\frac{i\hbar v V_0}{2} & 0 \\
-iE_{-1} & i\hbar vk_y & 0 & \frac{i\hbar v V_0}{2} \\
-\frac{i\hbar v V_0}{2} & 0 & -i\hbar vk_y & iE_0 \\
0 & \frac{i\hbar v V_0}{2} & -i E_0 & i\hbar vk_y \end{pmatrix} \begin{pmatrix}
\al_{-1} \\ \beta_{-1} \\ \al_0 \\ \beta_0 \end{pmatrix} ~=~ \hbar vk'_x
\begin{pmatrix}
\al_{-1} \\ \beta_{-1} \\ \al_0 \\ \beta_0 \end{pmatrix}. \eeq
(We have taken $m_0 = 0$, hence $k'_y = k_y$).
We then find that the values of $k'_x$ are given by
\beq \hbar^2 v^2 k_x^{'2} ~=~ \frac{1}{2} ~\Bigl[ ~E_0^2 +E_{-1}^2 -
\frac{\hbar^2 v^2 V_0^{2}}{2}- 2\hbar^2 v^2k_y^2 ~\pm~ \sqrt{(E_0^2-E_{-1}^2)^2
-\hbar^2 v^2 V_0^2 (E_{-1}-E_0)^2 +4 \hbar^4 v^4 V_0^2k_y^2} \Bigr].
\label{kx2} \eeq
Assuming that $V_0$ is small, we see that if $E_0$ is not equal to $\hbar
\om /2$, the change in $k_x^{'2}$ is of the order of $V_0^2$, but if $E_0 =
\hbar \om /2$, we get
\beq k_x^{'2} ~=~ k_{x,0}^2 ~\pm~ iV_0 k_{x,0}, \label{kx3} \eeq
giving a change in $k_x^{'2}$ of the order of $V_0$. Thus, for $V_0$ small,
$E_0 = \hbar \om /2$ gives a much larger change in $k'_x$ as compared to $E_0
\ne \hbar \om /2$. Next, Eq.~\eqref{kx3} implies that $k'_x = k_{x,0} \pm i
V_0 /2$ has an imaginary part given by $i V_0 /2$. Hence the wave function
$e^{i k'_x L}$ decays exponentially as $e^{-V_0 L/2}$ as we go across the 
barrier region from $x=0$ to $L$; this reduces the transmitted particle 
current and hence the conductance. We can find the width of the region of low 
conductance around $E_0 = \hbar \om /2$ by using Eq.~\eqref{kx2} to determine 
when $k'_x$ becomes complex. We find that this happens when
\beq |E_0 ~-~ \frac{\hbar \om}{2}| ~<~ \dfrac{\hbar v V_0}{2} ~\sqrt{1 ~-~
\frac{4v^2 k_y^2}{\om^2}}, \eeq
which is proportional to $V_0$. Thus the width of the dip in the conductance
is expected to be proportional to $V_0$ while the magnitude of the dip (which
is related to the transmission probability) should be proportional to
$e^{-V_0 L}$ since the transmission amplitude is proportional to
$e^{-V_0 L/2}$. If we hold $V_0 L$ fixed and vary $L$, the magnitude of the
dip should remain the same but the width should be proportional to $V_0$
and therefore inversely proportional to $L$. This agrees with the plot shown
in Fig.~\ref{fig17} where we have plotted the differential conductance as a
function of $E_0$ for two values of $L$, with $m_0 = 1$, $\om = 25.1$,
and $V_0 L = 2.3$. Thus by varying $E_0$, $\om$ and $V_0$, we can tune the
conductance and achieve a switching behavior close to $E_0= \hbar \om /2$.

\section{Discussion}
\label{sec5}

In this work, we have studied transport across a magnetic barrier
placed on the top surface of a three-dimensional TI, where the
barrier strength varies sinusoidally with time in addition to having
a constant term. Such a situation arises if a ferromagnetic strip is
placed on top of the TI surface and a periodically driven magnetic
field is applied to the ferromagnet; then the magnetization of the
ferromagnet would oscillate in time. If the magnetization points
along the $\hat x$ direction and has a Zeeman coupling to the spin
of the electrons on the TI surface, it would lead to the
time-dependent Hamiltonian that we have studied in this paper.
Alternatively, the ferromagnetic strip could have a time-independent
magnetization along the $\hat x$ direction, and a electromagnetic
field (which is linearly polarized along the $\hat y$ direction) can
be applied to the same region of the surface of the TI. The vector
potential of the electromagnetic field would point along the $\hat
y$ direction and vary periodically in time; this would lead to the
same time-dependent Hamiltonian.

We have studied two kinds of time-dependent magnetic barriers: a
$\de$-function barrier and a barrier which has a finite width $L$. The
$\de$-function barrier leads to a discontinuity of the wave function of a
specific kind. For the finite width barrier, we have to match the wave
function at the two edges of the barrier. In both cases, we have to take into
account a large number of Floquet bands. We consider what happens when an
electron is incident on the barrier from the left with an energy $E_0$ and an
angle of incidence $\ta$. We numerically calculate the transmitted current
as a function of $\ta$; integrating
this over $\ta$ gives the differential conductance $G$ when a voltage
equal to $E_0$ is applied to the leads. We have studied these quantities
as a function of $E_0$, the driving frequency $\om$, and the constant $m_0$
and the oscillation amplitude $V_0$ of the barrier strength.

Our main results are as follows. For the transmitted current, we
find kinks at certain values of $\ta$. These arise because the
momentum in the direction of propagation changes from a real to a
complex value in some side band at those values of $\ta$; the
corresponding wave function changes from a plain wave to an
exponentially decaying wave which then leads to a drop in the
transmitted current. For small values of $\om$ (the adiabatic
limit), one can approximate the results well by averaging over a
sequence of time-independent values of the barrier strength. We then
find that there are peaks in the conductance when $m_0$ is equal to
$\pm V_0$ since the barrier strength stays close to zero for long
periods of time. For large values of $\om$, we find numerically that
the transmission is dominated by three Floquet bands; the central
band $n=0$ and the first two side bands $n = \pm 1$. This allows us
to calculate the conductance more easily. We have shown that the
time-dependent barrier problem can be mapped to a time-independent
magnetic barrier with an effective strength $m_{eff}$ whose value
depends on the driving parameters. Thus the conductance of the
driven system can be effectively characterized by a static parameter.

Next, we have analyzed surface plots of the conductance as a
function of $V_0$ and $m_0$, for different values of $\om$. When
$\om$ and $V_0$ are large compared to $E_0$, and $m_0$ is small, we
find that the conductance has peaks at a discrete set of values of
$V_0$. To understand this better, we have made a detailed study of
the conductance as a function of $V_0$ and $\om$, keeping $E_0$ and
$L$ fixed and taking $m_0 = 0$. We find that the behavior of $G$ is
quite different depending on whether $\om /V_0$ lies above or below
a value equal to $0.83$ (this is related to the first zero of the
Bessel function $J_0 (z)$). When $\om /V_0 \gtrsim 0.83$, the
conductance is large along certain curves in the $(V_0,\om)$ plane;
these curves correspond to resonances, and their spacing is
proportional to $1/L$. When $\om /V_0 \lesssim 0.83$, the
conductance is generally small; however it is particularly small
along certain straight lines whose slopes are related to the
successive zeros of $J_0 (z)$. We present a Floquet perturbation
theory which can explain both the resonances and lines of very small
conductances. We have then studied the conductance as a function of
the incident energy $E_0$. We find that this shows a prominent dip
when $E_0$ is close to $\hbar \om/2$ when these quantities are much
larger than $m_0$ and $V_0$. We can understand this as follows. When
$E_0 = \hbar \om/2$, the energy in one of the side bands, $E_{-1}$,
becomes equal in magnitude to $E_0$. We then find numerically that
the conductance is dominated by only the two bands, $n=0$ and $-1$.
This allows us to analytically estimate the width and magnitude of
the conductance dips. We find that the width is proportional to
$V_0$, in agreement with the numerical results.

Our results can be experimentally tested as follows. As mentioned in
Sec.~\ref{sec2b}, an energy of $0.01$ eV corresponds to a frequency
scale of about 15 THz. This gives an estimate of the required
frequency $\om$ if electromagnetic radiation is used to produce the
time-dependent magnetic barrier. The strength $V_0$ of the magnetic
barrier is proportional to the vector potential $\cal A$; this is
related to the electric field $\cal E$ (as ${\cal E} = \om {\cal
A}$) and hence to the intensity of the radiation. Our most important
result is that when $\om$ is much larger than the energy $E_0$ of
the electrons ($\om \gg E_0/\hbar$), the conductance across the
barrier has a striking dependence on $V_0$ and $\om$ as indicated in
Fig.~\ref{fig09}. For instance, increasing the intensity of the
radiation keeping the frequency fixed or decreasing the frequency
keeping the intensity fixed should reduce the conductance sharply as
we cross a line in the $(V_0,\om)$ plane. Further, the conductance
shows resonance-like features when the conductance is large, and
these features depend sensitively on the width of the barrier.
Another interesting feature appears when $\om$ is of the order of
$E_0/\hbar$; namely, there is a prominent dip in the conductance
when $\om$ crosses $2 E_0/ \hbar$, and this dip becomes sharper as
the barrier length is increased.

Finally, we have used a semiclassical approach to study this problem
in the limit where the spin of the particle is very large instead of
being 1/2. This allows us to use classical equations of motion to
study the motion of the particle in the presence of a time-dependent
magnetic barrier, assuming that at time $t=0$ it is incident on the
left edge of the barrier at a normal angle of incidence. We find
that although the motion of the particle can be quite complicated
inside the barrier, it eventually escapes either to the left or to
the right of the barrier. To connect this with the study of spin-1/2
electrons done in the earlier sections, we interpret escaping to the
left (right) as reflection (transmission) and therefore as small and
large conductances respectively. With this interpretation, we find
that the behaviors of the system as a function of the barrier
parameters $\om$, $m_0$ and $V_0$ show a qualitative match between
spin-1/2 and the large spin limit.

We have not considered the effects of disorder and have only studied
ballistic transport in this work. If there is strong disorder, the
mean free path of the electrons becomes less than the width of the
magnetic barrier, and the effect of disorder would have to be considered.
This would be an interesting problem to study in the future. However, our
results should hold with minor modifications in the weak disordered
limit~\cite{fogler}. Moreover, we have not addressed the effects of
interactions between the Dirac electrons which
may lead to heating effects, specially at low drive frequencies. A
detailed study of this problem is left as a subject of future study.

\vspace{.8cm}
\centerline{\bf Acknowledgments}
\vspace{.5cm}

D.S. thanks Amit Agarwal, Sankalpa Ghosh and Puja Mondal for stimulating
discussions. D.S. also thanks DST, India for Project No. SR/S2/JCB-44/2010
for financial support.

\vspace{.5cm}

\appendix

\section{Basics of Floquet theory}
\label{basics}

In this section we briefly recapitulate Floquet
theory~\cite{bukov,mikami}. Given a Hamiltonian which varies
periodically in time with a time period $T = 2\pi/\om$, namely,
$H(t+T)= H(t)$, we want to find the solutions of the equation $i
\hbar \pa \psi(t) /\pa t = H(t) \psi (t)$. To this end, we define
the Floquet operator which time evolves the system through one time period,
\beq U ~=~ {\cal T}~ \exp [- ~\frac{i}{\hbar} ~\int_0^T ~dt H(t)],
\label{flo1} \eeq
where $\cal T$ denotes time-ordering. Thus,
$\psi (T) = U \psi (0)$. Since $U$ is a unitary operator, its
eigenvalues must be phases; denoting the $n$-th eigenvalue and
eigenstate as $e^{-i\ta_n T}$ and $\psi_n (0)$, we have \beq U \psi_n
(0) ~=~ e^{- i \ta_n T} ~\psi_n (0). \label{flo2} \eeq We can find
$\ta_n$ and $\psi_n (0)$ as follows. Eq.~\eqref{flo2} implies that
$\psi_n (T) = e^{- i \ta_n T} \psi_n (0)$. We can therefore write \beq
\psi_n (t) ~=~ e^{- i \ta_n t} ~\sum_{m=-\infty}^\infty ~e^{-im \om
t} \psi_{n,m}, \label{psit} \eeq so that $\psi_n (0) =
\sum_{m=-\infty}^\infty \psi_{n,m}$. Next, the periodicity of the
Hamiltonian in time means that we can write \beq H(t) ~=~
\sum_{m=-\infty}^\infty ~H_m ~e^{-im \om t}, \label{hamt1} \eeq where
$H_m$ is given by $H_m = (1/T) \int_0^T dt H(t) e^{i m \om t}$. The
equation $i \hbar \pa \psi(t) /\pa t = H(t) \psi (t)$ then leads to
the infinite set of coupled time-independent equations, 
\beq \sum_{p=-\infty}^\infty ~(H_{m-p} ~-~ m \hbar \om ~\de_{m,p}) ~
\psi_{n,p} ~=~ \hbar \ta_n~ \psi_{n,m}. \label{psinm} \eeq 
where $m$ runs over all values from $- \infty$ to $\infty$. We can solve 
these equations to find $\ta_n$ and $\psi_n (0)$; a numerical solution 
typically requires an upper and a lower cut-off for the values of $p$.

\section{Semiclassical approach}
\label{semiclassical}

In the earlier sections, we have studied the transmission of spin-1/2
electrons through a time-dependent magnetic barrier. It may be instructive,
however, to look at the more general problem of the transmission of a spin-$S$
particle whose spin angular momentum operators $\vec S$ satisfy ${\vec S}^2
= S (S+1) \hbar^2$. In particular, we will study the semiclassical limit
$S \to \infty$ and see if this can give a qualitative understanding of some
of the phenomena that we have discovered for the spin-1/2 case.

In this section, we investigate the semiclassical dynamics of a charged
particle with spin~\cite{maiti} which is moving in two dimensions in the
presence of a time-dependent magnetic barrier with a finite width. The
dynamics involves two two-dimensional vectors, namely, the position of the
particle ${\vec r} = (x,y)$ and the canonically conjugate momentum ${\vec p}
= (p_x, p_y)$. In addition, the dynamics also involves a three-dimensional
unit vector ${\vec n} = (n_1, n_2, n_3)$ which is related to the spin vector
$\vec S$ as ${\vec n} = {\vec S}/(S \hbar)$; in the limit $S \to \infty$,
$\vec n$ becomes a unit vector. For convenience, we will denote $M = S \hbar$.

Classically, the different dynamical variables describing the particle
satisfy the following Poisson bracket relations
\beq \{ r_i, p_j \}_{PB} ~=~ \de_{ij} ~~~{\rm and}~~~ \{n_i, n_j \}_{PB} ~=~
\frac{1}{M} ~\sum_{k=x,y,z} \ep_{ijk} ~n_k, \label{pb} \eeq
where $\ep_{ijk}$ is the totally antisymmetric tensor with $\ep_{123} = 1$.
All other Poisson brackets such as $\{ r_i, n_j \}$ and $\{ p_i, n_j \}$
vanish. We now consider a Hamiltonian of the form~\cite{maiti}
\bea H &=& v \vec{n} \cdot \vec{\Pi} ~+~ q \phi, \non \\
{\vec \Pi} &=& {\vec p} ~-~ q {\vec A}, \label{ham3} \eea
where $\vec A$ is the vector potential, and $\phi$ is the electrostatic
potential. [We note that in two dimensions, the electric field ${\vec E} =
-{\vec \nabla} \phi - \pa {\vec A}/{\pa t}$ has two components and the
magnetic field $B = \pa A_y /\pa x - \pa A_x /\pa y$ has one component].
Using Eq.~\eqref{pb} we find that $\{ \Pi_x, \Pi_y \}_{PB} = B$.
The Hamiltonian that we have studied earlier, given in Eq.~\eqref{psi2},
has the form $v {\vec \sigma} \times {\vec \Pi}$ while the Hamiltonian
given in Eq.~\eqref{ham3} has the form $v \vec{n} \cdot \vec{\Pi}$. The two
are related by first rotating by $\pi/2$ which transforms $\si^x \to \si^y$
and $\si^y \to - \si^x$, and then going from spin-1/2 to the large spin limit.

Now, the classical equations of motion of a dynamical variable $\cal O$ is
given by
\beq \frac{d {\cal O}}{dt} ~=~ \frac{\pa {\cal O}}{\pa t} ~-~ \{ H,
{\cal O} \}_{PB}. \eeq
Using Eqs.~\eqref{pb}, we obtain the following equations
\bea \dot{x} &=& v n_1, ~~~~~~ \dot{y} ~=~ v n_2, \non \\
\dot{\Pi}_x &=& q ~[E_x ~+~ v n_2 B], ~~~~~~ \dot{\Pi}_y ~=~ q ~[E_y ~-~
v n_1 B], \non \\
\dot{n}_1 &=& \frac{v}{M} ~n_3 \Pi_y, ~~~~~~ \dot{n}_2 ~=~ - ~\frac{v}{M} ~
n_3 \Pi_x, ~~~~~~ \dot{n}_3 ~=~ \frac{v}{M} ~(n_2 \Pi_x ~-~ n_1 \Pi_y).
\label{eom1} \eea
Note that these equations preserve the constraint ${\vec n}^2 = 1$.

Before proceeding further, we present a simple case of Eqs.~\eqref{eom1}.
In the absence of electromagnetic fields, i.e., for a free particle,
we find that Eqs.~\eqref{eom1} are invariant under rotations in the $x-y$
plane. We then find that after a suitable rotation, the solution of the
equations can be written as
\bea \Pi_x &=& p, ~~~~~~ \Pi_y ~=~ 0, \non \\
n_1 &=& \mu, ~~~~~~ n_2 ~=~ \sqrt{1 ~-~ \mu^2} ~\cos (\frac{vpt}{M} ~+~ \al),
~~~~~~ n_3 ~=~ \sqrt{1 ~-~ \mu^2} ~\sin (\frac{vpt}{M} ~+~ \al), \non \\
x &=& v \mu t ~+~ x_0, ~~~~~~ y ~=~ \frac{M \sqrt{1 ~-~ \mu^2}}{p} ~\sin
(\frac{vpt}{M} ~+~ \al) ~+~ y_0, \eea
where $p, ~\mu, \al, x_0$ and $y_0$ are constants, and $\mu$ must lie
in the range $[-1,1]$. Note that the particle moves with constant velocity
$v \mu$ in one direction ($x$) but oscillates in the transverse direction
($y$). The velocity $v \mu$ can take any value from
$+v$ to $-v$. This agrees with the fact that if a spin-$S$ particle has
a Hamiltonian of the form $H = (v/\hbar S) {\vec S} \cdot {\vec p}$, the
group velocity is given by $v_g = (v/\hbar S) {\vec S} \cdot {\hat p}$.
Then the fact that eigenvalues of $(1/\hbar) {\vec S} \cdot {\hat p}$ are
quantized as $S, ~S-1, ~\cdots, ~- S$, implies that $v_g$ can take $2S+1$
values in the range $[-v,v]$. In the limit $S \to \infty$, $v_g$ can
take all values in the above range.

Returning to our problem described by Eq.~\eqref{psi2}, $\phi$ and
$A_x$ are equal to zero, while the term multiplying $v \si^x$ in the
barrier region is equal to $\hbar k_y + \hbar \{ m_0 + V_0 \cos (\om t) \}$.
To identify this with $\Pi_y = p_y - q A_y$, we must take
\beq q A_y ~=~ - ~\hbar \{ m_0 + V_0 \cos (\om t) \} ~~~{\rm for} ~~~ 0 < x <
L, \eeq
and zero for other values of $x$. However, such a step function form for
$A_y (x,t)$ would mean that the magnetic field $B$ would blow up as a
$\de$-function at $x=0$ and $L$. We will therefore approximate $A_y$ to
have the continuous and piecewise linear form
\bea q A_y (x,t) &=& 0 \quad {\rm for}~~~ x ~<~ 0, \non \\
&=& \dfrac{x}{\de} ~\{ m_0 ~+~ V_0 \cos (\om t)\} \quad {\rm for} ~~~
0 ~<~ x ~<~ \de, \non \\
&=& m_0 ~+~ V_0 \cos (\om t) \quad {\rm for}~~~ \de ~<~ x ~<~ L ~-~ \de,
\non \\
&=& \dfrac{L ~-~ x}{\de} ~\{ m_0 + V_0 \cos (\om t) \} \quad {\rm for} ~~~
L ~-~ \de ~<~ x ~<~ L, \non \\
&=& 0 \quad {\rm for} ~~~ x ~>~ L, \label{ay} \eea
where $\de$ is a small distance ($\de \ll L$) over which $A_y$ changes between
zero and the value that it has inside the barrier. We note that the parameters
$m_0, ~V_0$ in the above equations are not identical to the same parameters in
Eq.~\eqref{psi2}; they differ by a factor of $\hbar$.

We can now study the time evolution given by Eqs.~\eqref{eom1}, where
\beq q E_x ~=~ 0, ~~~~q E_y ~=~ - ~q ~\pa A_y /\pa t, ~~~~q B ~=~ q~ \pa
A_y /\pa x, \eeq
and $q A_y$ is given in Eq.~\eqref{ay}. We have to choose some initial
conditions at time $t=0$. We will assume that the particle comes in from
$x = - \infty$ with an energy $E > 0$ and arrives at $x=0, ~y=0$ at a normal
angle of incidence. Hence $x(0) = 0$ (i.e., the left edge of the barrier),
$y(0) = 0$, while $\Pi_x = p_x = E/v$ and
$\Pi_y = p_y - q A_y = 0$ at $t=0$. Finally, the spin-momentum locking implied
by the Hamiltonian in Eq.~\eqref{ham3} implies that since $\Pi_x (0) > 0$ and
$\Pi_y (0) = 0$, we must take choose the components of the unit vector
$\vec n$ as $n_1 (0) = 1$ and $n_2 (0) = n_3 (0) = 0$. To summarize, the
initial conditions are
\beq x(0) ~=~ 0, ~~~~~~ y(0) ~=~0, ~~~~~~ \Pi_x (0) ~=~ \frac{E}{v}, ~~~~~~
\Pi_y (0) ~=~ 0, ~~~~~~ n_1 (0) ~=~ 1, ~~~~~~ n_2 (0) ~=~ 0, ~~~~~~ n_3 (0)
~=~ 0. \label{init} \eeq
We can now use Eqs.~\eqref{eom1} to numerically find how the different
dynamical variables change with time. Since we are specifically interested in
transmission through the barrier, we will concentrate on the variable $x$.

\begin{figure}[H]
\centering
\includegraphics[scale=0.5]{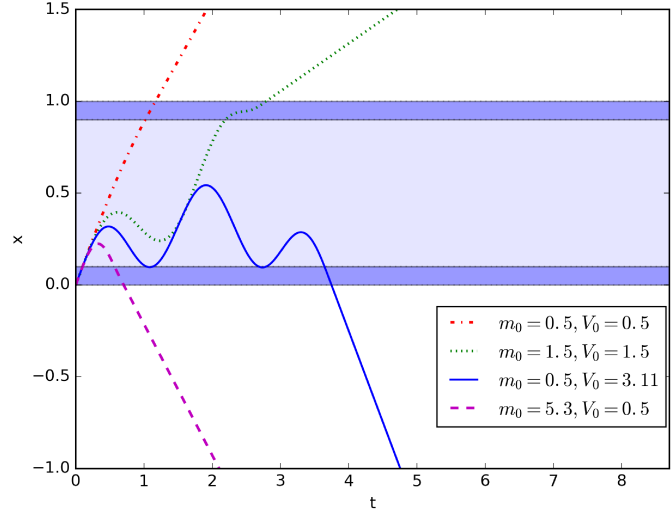} 
\caption{$x$ as a function of $t$ for $\om = 0.01$ and different values of
$m_0$ and $V_0$} \label{fig18} \end{figure}

\begin{figure}[H]
\centering
\includegraphics[scale=0.5]{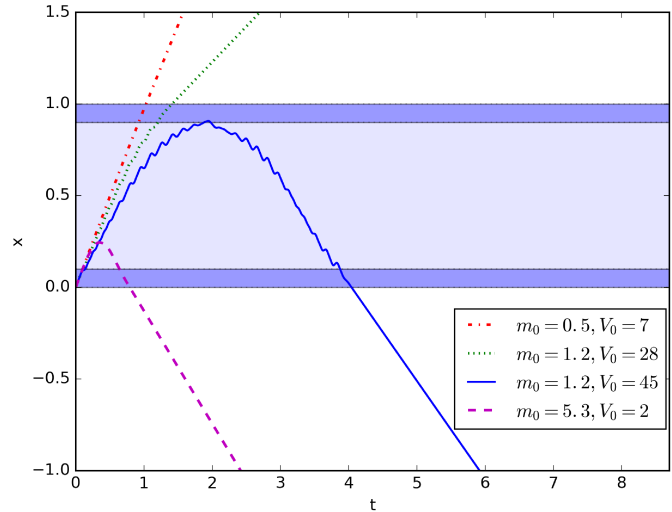} 
\caption{$x$ as a function of $t$ for $\om = 40.1$ and different values of
$m_0$ and $V_0$.} \label{fig19} \end{figure}

Figures~\ref{fig18} and \ref{fig19} show $x$ as a function of $t$ for
different values of $\om$, $m_0$ and $V_0$. We have taken $E = 2$ and $v=1$ in
Eq.~\eqref{init} and the barrier width $L=1$ and $\de = 0.1$ in appropriate
units. In these figures, we see that depending on the various parameters,
the particle may have a complicated trajectory while it is inside the
barrier region ($0 < x < L$), but eventually it always escapes
either to the left ($x < 0$) with constant negative velocity or to the
right ($x > L$) with a constant positive velocity. In order to compare
with the results obtained in the earlier sections, we can interpret escape
to the left as reflection (hence zero or small conductance) and to the
right as transmission (large conductance).


In Fig.~\ref{fig18}, we see that for small and intermediate equal values of
$m_0$ and $V_0$ (dash dot red and dotted green lines), there is transmission
which maps to the $m_0 -V_0$ plane in Fig.~\ref{fig08} (a). But as we increase
$m_0$ keeping $V_0$ fixed, or increase $V_0$ keeping $m_0$ fixed, we see in
Fig.~\ref{fig18} (solid blue and dashed magenta lines) that there is
reflection. This maps to regions in Fig.~\ref{fig18}, where there is no
conductance. Also in Fig.~\ref{fig18} (solid blue line), where we have fine
tuned the value of $V_0$, we see that the particle remains in the barrier
region for a relatively longer time.

In Fig.~\ref{fig19} we see that the value of $V_0 = 41.8$ is a cut-off, below
which there is transmission and above which there is reflection. For small
$m_0$, a comparison of Figs.~\ref{fig18} and \ref{fig19} shows that as $\om$
is increased, the value of $V_0$ beyond which there is reflection increases.
This maps qualitatively to the surface plot in Fig.~\ref{fig08} (c)
where we see that for larger $\om$, the value of $V_0$ up to which there is
large conductance increases. Although the cut-off values of $V_0$ do not
match between the surface plots and the semiclassical analysis, we see that
there is a qualitative mapping between the two as a function of $\om$.

%

\end{document}